\documentclass[pra,showpacs,preprintnumbers]{revtex4-2}
\usepackage{amsmath}
\usepackage{amssymb} 
\usepackage{graphicx}
\usepackage{bm}
\usepackage{comment}
\usepackage{mathrsfs} 
\usepackage{subeqnarray} 
\usepackage{subfigure} 
\usepackage{hyperref}
\usepackage{color}
\usepackage{fancyhdr}
\usepackage{datetime}
\numberwithin{equation}{section} 
\usepackage{ifpdf} 
\begin{document}
\ifpdf
 \def\anglefig{0}
\else
 \def\anglefig{-90}
\fi
\title{
Distribution of the total angular momentum in relativistic configurations
}
\author{Michel Poirier}
\email{michel.poirier@cea.fr}
\affiliation{CEA - Paris-Saclay University, Laboratory ``Interactions, Dynamics, 
and Lasers'', CE Saclay, F-91191 Gif-sur-Yvette, France}
\author{Jean-Christophe Pain}
\email{jean-christophe.pain@cea.fr}
\affiliation{CEA, DAM, DIF, F-91297 Arpajon, France}
\affiliation{Universit\'e Paris-Saclay, CEA, Laboratoire Mati\`ere en Conditions 
 Extr\^emes, F-91680 Bruy\`eres-le-Ch\^atel, France}
\date{\today}

\begin{abstract}
This paper is devoted to the analysis of the distribution of the total angular 
momentum in a relativistic configuration. Using cumulants and generating 
function formalism this analysis can be reduced to the study of individual
subshells with $N$ equivalent electrons of momentum $j$.
An expression as a nth-derivative is provided for the generating function of the 
$J$ distribution and efficient recurrence relations are established. 
It is shown that this distribution may be represented by a Gram-Charlier-like 
series which is derived from the corresponding series for the magnetic quantum 
number distribution. The numerical efficiency of this expansion is fair when 
the configuration consists of several subshells, while the accuracy is less good 
when only one subshell is involved. An analytical expression is given for the 
odd-order momenta while the even-order ones are expressed as a series which provides 
an acceptable accuracy though being not convergent. Such expressions may be 
used to obtain approximate values for the number of transitions in a spin-orbit 
split array: it is shown that the approximation is often efficient when few terms 
are kept, while some complex cases require to include a large number of terms.
\end{abstract}

\maketitle

\section{Introduction}

In order to model the emission and absorption spectral properties 
of hot plasmas, in the context of stellar physics or in laser-plasma experiments, 
for inertial-confinement-fusion studies for instance,  one has to properly describe 
multi-electron configurations with several open subshells. In particular, knowing 
\emph{a priori} the number of lines between  two configurations is of great interest. 
The statistics of electric-dipole (E1) lines was investigated by Moszkowski 
\cite{Moszkowski1960}, Bauche and Bauche-Arnoult \cite{Bauche1987}, and more recently 
by Gilleron and Pain \cite{Gilleron2009}. 
The number of lines is a cornerstone of opacity codes, in order to decide whether
a transition array can be modeled statistically, using the above-mentioned methods, 
or needs a detailed-line accounting calculation, involving the diagonalization of 
the Hamiltonian \cite{Porcherot2011,Pain2015}. Alternative methods such as the 
partially-resolved-transition-array approach 
\cite{Iglesias2012,Iglesias2012a,Iglesias2012b}, and its extension to the 
superconfiguration formalism \cite{Wilson2015,Kurzweil2016}, can be applied when the 
number of lines of a transition array exceeds a particular value. The statistics of 
electric quadrupole (E2) lines was also studied \cite{Pain2012}. 

In counting problems, the generating-function technique is a powerful tool, 
whether to obtain analytical expressions, to derive recursion relations, 
or to find approximate formulas. 
The generating function leads also to the determination of cumulants, which are 
an important ingredient of statistical modeling and from which the moments may be  
obtained. 
In this framework, we recently published analytical formulas and 
recurrence relations for the number of electronic configurations in a 
superconfiguration \cite{Pain2020}, together with a statistical analysis based 
on the calculation of cumulants.

The determination of the total angular momentum multiplicities was first 
investigated by nuclear physicists \cite{Bethe1936} in the framework of the 
shell model \cite{Elliott1957,Moszkowski1957} and later by atomic physicists for 
electronic configurations. Considering a system of $N$ identical fermions, the 
problem boils down to deducing the allowed total angular momenta $J$ to which they 
may couple. Some values of $J$ are forbidden by anti-symmetrization due to the Pauli 
exclusion principle, some others occur more than once. 
As pointed out by Condon and Shortley, the number $Q(J)$ of levels with 
angular momentum $J$ is equal to the number of states with projection $M=J$ minus 
the number of states with $M=J+1$.
In his 1936 paper \cite{Bethe1936}, Bethe modeled the distribution of angular 
momentum by the Wigner-type distribution which can be deduced from a Gaussian form 
of the distribution $P(M)$. Years later, Bauche and Bauche-Arnoult derived 
analytical formulas for $Q(J)$ and for the number of lines between two electronic 
configurations, assuming a fourth-order Gram-Charlier expansion series of $P(M)$ 
\cite{Bauche1987}. At the same period, Hirst and Wybourne used statistical group 
theory to determine the parameters of the Wigner-type approximate formula and 
pointed out a connection with the theory of partitions of integers 
\cite{Hirst1986}. 

In a previous work \cite{Poirier2021}, cited here as paper I, 
we proposed a statistical analysis of the distribution $P(M)$ in the case 
of a relativistic configuration (i.e., made of subshells $j^N$). Using the 
generating-function formalism and the properties of Gaussian binomial 
coefficients, we derived recursion relations for $P(M)$ as well as analytical 
expressions for the cumulants. In the same work, we carried out an analysis 
of the distribution $P(M)$ using Gram-Charlier and Edgeworth expansion series 
at any order. 

However, although $Q(J)$ can be deduced from $P(M)$, only very few properties 
of $Q(J)$ are explicitly known. 
The aim of the present paper is multiple. First, following our analysis of 
the magnetic quantum number distribution, we propose here an expression 
for the generating function of the distribution $Q(J)$, from which we derive 
efficient recurrence relations. Second, using paper I analysis, we generalize 
the above-mentioned Bethe formula in the form of a Gram-Charlier-like 
series the convergence of which is analyzed. Third, we propose expressions 
for the moments $\sum_J (2J+1)^nQ(J)$, in a closed analytical form for odd 
$n$, and as a cumulant-based expansion for even $n$, with a study of its 
convergence. This sheds some light on the delicate question about the number 
of levels $\sum_J Q(J)$ inside a given relativistic configuration. 
Another useful application of this formalism is that it allows one 
to derive the number of lines inside a transition array, which is a 
significant information when analyzing the complex spectra such as those 
obtained in high-temperature plasmas.

\section{Generating function and recurrences}\label{sec:genfunc}
\subsection{Population distribution of the total angular momentum}
In the most general relativistic configuration, the determination of the 
distribution of the total angular momentum $Q(J)$ is connected to the 
distribution of the magnetic quantum number $P(M)$ through the relation 
\cite{Bauche1987}
\begin{subequations}\begin{gather}
Q(J_\text{max}) = P(J_\text{max})\\
Q(J) = P(M=J)-P(M=J+1)\label{eq:QJvsPM}
\end{gather}\end{subequations}
where in the latter equation one has $0 \le J < J_\text{max}$. However for 
the sake of analyticity, it is very important to allow for \emph{any} value 
of $J$ between $-J_\text{max}$ and $J_\text{max}-1$ in Eq.(\ref{eq:QJvsPM}), 
keeping in mind that the ``unphysical'' $Q(J)$ are convenient mathematical 
tools. Using the symmetry property $P(-M)=P(M)$ one easily verifies from the 
property (\ref{eq:QJvsPM}) imposed on any $J$ that 
\begin{equation}
Q(-J-1)=P(-J-1)-P(-J)=-(P(J)-P(J+1))=-Q(J)\label{eq:symQ}
\end{equation}
for any $J\le J_\text{max}$. This means that $Q(J)$ is an odd function 
versus the origin $J_0=-1/2$. In other words if $N$ is odd $Q(-1/2)=0$, 
$Q(-3/2)=-Q(1/2)$ \dots, if $N$ is even $Q(-1)=-Q(0)$, $Q(-2)=-Q(1)$ \dots
Finally for $J < -J_\text{max}-1$ or $J > J_\text{max}$, $Q(J)$ vanishes. 
One must notice that from such a definition, one has 
\begin{equation}
\sum_{J=-J_\text{max}-1}^{J_\text{max}}Q(J)=0\label{eq:sumQ}
\end{equation}
which prevents us from normalizing the $Q(J)$ distribution.

\subsection{Analytical expression of the generating function and recurrences}
In this subsection we consider relativistic configurations consisting 
of a single subshell with $N$ equivalent electrons $j^N$. 
In paper I, we have shown that the distribution $P(M)$ 
can be obtained from the generating function --- 
$z$ being the arbitrary complex variable of this function ---
\begin{equation}\label{eq:FjNz}
\sum_M P(M)z^M =z^{-J_\text{max}}\prod_{p=1}^N\frac{z^{2j+2-p}-1}{z^p-1}
 =z^{-J_\text{max}}\mathscr{F}(j,N;z)
\end{equation}
where $\mathscr{F}(j,N;z)$ is the product $\prod_{p=1}^N
(z^{2j+2-p}-1)/(z^p-1)$ also known as the Gaussian binomial coefficient.
From the general property (\ref{eq:QJvsPM}) of the distribution $Q$ and 
from the above relation (\ref{eq:FjNz}) one gets at once
\begin{equation}
z^{J_\text{max}+1}\sum_J Q(J)z^J 
 = z^{J_\text{max}+1} \sum_J\left(P(J)-P(J+1)\right)z^J
 = (z-1) z^{J_\text{max}}\sum_M P(M)z^M\label{eq:sumQJzJ}
\end{equation}
which proves that the generating function for $Q(J)$ is simply equal to the 
generating function for $P(M)$ multiplied by $(z-1)$. One will note that this 
property holds for any relativistic configuration, containing one or several 
subshells. 

This important results bears a series of consequences. The first one is that 
the numbers $Q(J)$ may be obtained as a simple nth-derivative when a single 
subshell $j^N$ is involved. Similarly to Eq. (\ref{eq:FjNz}) one has
\begin{equation}\label{eq:GjNz}
\mathscr{G}(j,N;z)= \sum_{n=0}^{2J_\text{max}+1} Q(n-J_\text{max}-1)z^n 
 = (z-1)\prod_{p=1}^N\frac{z^{2j+2-p}-1}{z^p-1}.
\end{equation}
This allows one to express the population $Q(J)$ as a multiple derivative 
\begin{equation}\label{eq:QJ_derivn}
Q(n-J_\text{max}-1) = \frac{1}{n!}\left.\frac{d^n}{dz^n}\left[(z-1)
 \prod_{p=1}^N\frac{z^{2j+2-p}-1}{z^p-1}\right] \right|_{z=0}
\end{equation}
where $n$ varies from 0 to $2J_\text{max}+1$. From the symmetry property 
(\ref{eq:symQ}) one has also 
\begin{equation}Q(n-J_\text{max}-1) = -Q(J_\text{max}-n)\end{equation} 
which allows one to obtain $Q(J)$ for positive $J$ with less derivative 
operations.

Recurrence relations can be derived in a similar manner as in paper I. 
One may use the identity straightforwardly derived from the generating 
function (\ref{eq:GjNz})
\begin{equation}
 \mathscr{G}(j,N;z)(z^N-1)=\mathscr{G}(j,N-1;z)(z^{2j+2-N}-1)
\end{equation}
and derive $n$ times using the Leibniz rule. One may also use the relation 
on the $M$ distribution obtained in paper I (with some rewriting)
\begin{equation}
 P(M-N;j,N)-P(M;j,N) = P(M-j-1;j,N-1)-P(M+j+1-N;j,N-1)
\end{equation}
--- the indices $j,N$ are added when necessary for a correct understanding ---
and the fundamental relation (\ref{eq:QJvsPM}), which gives the 
recurrence property
\begin{equation}\label{eq:recQJ_overN}
 Q(J-N;j,N)-Q(J;j,N) = Q(J-j-1;j,N-1)-Q(J+j+1-N;j,N-1).
\end{equation}
With the definition 
\begin{equation}\mathscr{Q}_{j,N}(n)=Q(n-N(2j+1-N)/2-1;j,N)\end{equation}
one writes the somewhat simpler formula
\begin{equation}
\mathscr{Q}_{j,N}(n-N)-\mathscr{Q}_{j,N}(n)
 = \mathscr{Q}_{j,N-1}(n-2j-2+N)-\mathscr{Q}_{j,N-1}(n).
\end{equation}
The recurrence is initialized by the $N=1$ value
\begin{equation}Q(J;j,N=1)=\delta_{J,j}-\delta_{J,-j-1}.\end{equation}
In order to avoid the consideration of the full range of $n$ values one 
may use the symmetry property (\ref{eq:symQ}). 

Accordingly, a recurrence relation on $j$ is obtained by changing $j$ by 
1/2 while the number of electrons $N$ is kept constant. From
\begin{equation}
 \mathscr{G}(j+1/2,N;z)(z^{2j+2-N}-1)=\mathscr{G}(j,N;z)(z^{2j+2}-1)
\end{equation}
and multiple derivation using Leibniz rule, or from the paper I relation
\begin{equation}
 P(M-2j-1+N;j,N)-P(M;j,N) = P\left(M-2j-1+\frac{N}{2};j-\frac12,N\right)
  -P\left(M+\frac{N}{2};j-\frac12,N\right),
\end{equation}
one gets the second recurrence relation
\begin{equation}\label{eq:recQJ_overj}
 Q(J-2j-1+N;j,N)-Q(J;j,N) = Q\left(J-2j-1+\frac{N}{2};j-\frac12,N\right)
 -Q\left(J+\frac{N}{2};j-\frac12,N\right).
\end{equation}
With the $\mathscr{Q}_{j,N}$ quantities one has
\begin{equation}
\mathscr{Q}_{j,N}(n-2j-1+N) - \mathscr{Q}_{j,N}(n)
 = \mathscr{Q}_{j-1/2,N}(n-2j-1) - \mathscr{Q}_{j-1/2,N}(n).
\end{equation}
The recurrence is initialized using
\begin{equation}Q(J;j=0,N)=\delta_{J,0}-\delta_{J,-1}.\end{equation}

One notes that the same recurrence relations formally hold for $P(M)$ 
and $Q(J)$. Of course the solutions for $P(M)$ and $Q(J)$ differ because 
the initial values ($N=1$ or $j=0$) are different for $P$ and $Q$. 
It has been shown in paper I that such relations allow one to get 
the full set of $P(M)$ with a minimum of computations. Of course the 
same statement holds for the derivation of the $Q(J)$ distribution.

\subsection{Application to the analysis of even-odd staggering in 
angular-momentum distribution}
It has been shown by Bauche and Coss\'e \cite{Bauche1997} and Pain 
\cite{Pain2013} that when the number of electrons $N$ in a subshell is even, 
the levels with even angular momentum $\sum_{J\text{ even}} Q(J)$ 
outnumber the levels with odd angular momentum $\sum_{J\text{ odd}}Q(J)$.
This effect was proven to exist for non-relativistic as well as relativistic 
configurations. Expressions were derived in Ref.~\cite{Bauche1997} for the 
excess $\sum_{J\text{ even}}Q(J)-\sum_{J\text{ odd}} Q(J)$ 
by an explicit counting of the levels. As shown in Ref.~\cite{Pain2013} 
and in Appendix \ref{sec:staggering} the present generating-function 
formalism allows for a simpler derivation of the excess evaluation by 
computing $\mathscr{G}(j;N;-1)$. 
Moreover, other sum rules may also be obtained by computing this 
generating function at other points on the trigonometric circle.

\section{Approximation by Gram-Charlier-like series}\label{sec:psGC} 
\subsection{Analytical formulation}
The generating function for the $J$-momentum distribution (\ref{eq:GjNz}) 
being straightforwardly derived from the generating function (\ref{eq:FjNz}) 
for the magnetic quantum number, one might expect a similar property to hold 
when transposing the cumulant generating function 
to the $Q(J)$ distribution. However one must note that the standard 
definition for this generating function $K(t)$ 
\begin{equation}
 \exp(K(t)) = \left<\exp(tM)\right>
 = \sum_M P(M)e^{tM} \left/ \sum_M P(M) \right.\label{eq:defeKt}
\end{equation}
expresses it as the logarithm of a sum, and that this sum is normalized 
by $\sum_M P(M)$. As mentioned above (\ref{eq:sumQ}) the sum of the $Q(J)$ 
vanishes. Therefore a standard Gram-Charlier analysis is precluded here. 
A workaround consists in using the Gram-Charlier series for the $P(M)$ 
distribution and evaluating the difference $P(J)-P(J+1)$. In paper I we 
obtained the expansion
\begin{subequations}\begin{align}\label{eq:GCQJ_sum}
  P_\text{GC}(M)&=\frac{G}{(2\pi\sigma^2)^{1/2}}
   \exp\left(-\frac{M^2}{2\sigma^2}\right) 
   \sum_{n} c_{2n}He_{2n}\left(\frac{M}{\sigma}\right)
\end{align}
where $G$ is the degeneracy factor, which is, for the most 
general relativistic configuration $j_1^{N_1}\cdots j_w^{N_w}$
\begin{equation}\label{eq:norm_PM}
 G=\prod_{s=1}^{w}\binom{2j_s+1}{N_s}
\end{equation}
and $\sigma$ is the standard deviation for the $M$-distribution 
\begin{equation}\label{eq:sigma2_PM}
 \sigma^2=\sum_{s=1}^{w}\frac{j_s+1}{6}N_s(2j_s+1-N_s).
\end{equation}\end{subequations}
In Eq. (\ref{eq:GCQJ_sum}) $He_{2n}(X)$ is the Chebyshev-Hermite 
polynomial \cite{Abramowitz1972}
\begin{equation}He_{N}(X) =
 (-1)^Ne^{X^2/2}\frac{d^N}{dX^N}\left(e^{-X^2/2}\right)
 = N!\sum_m\frac{(-1)^mX^{N-2m}}{2^m m!(N-2m)!}.\label{eq:Hermite}
\end{equation}
The expression for the coefficients $c_{2n}$ is detailed below. 
The distribution for $J$ is therefore given by the finite difference 
formula, assuming the series is convergent
\begin{equation}\label{eq:diff_f}
Q_\text{GC}(J) = P_\text{GC}(J)-P_\text{GC}(J+1) = 
-\sum_{p=0}^\infty \frac{2^{-2p} }{(2p+1)!}
\frac{\partial^{2p+1}}{\partial J^{2p+1}}P_\text{GC}(J+1/2).
\end{equation}
The multiple derivative in the above formula is obtained from 
the explicit expression (\ref{eq:GCQJ_sum}) and from the property 
of Hermite polynomials, easily obtained from the above relation 
(\ref{eq:Hermite}), 
\begin{equation}\label{eq:derp_He}
\frac{d^p}{dX^p}\left(e^{-X^2/2}He_n(X)\right) = 
(-1)^p e^{-X^2/2}He_{n+p}(X).
\end{equation}
Putting such expression in the expansion (\ref{eq:GCQJ_sum}) we get
\begin{subequations}\begin{align}
 Q_\text{GC}(J) &= P_\text{GC}(J)-P_\text{GC}(J+1)\\
 &= \frac{G}{(2\pi)^{1/2}\sigma^2}
  \exp\left(-\frac{(J+1/2)^2}{2\sigma^2}\right)
  \sum_{p=0}^{\infty}\frac{(2\sigma)^{-2p}}{(2p+1)!}
  \sum_{n=0}^{\infty}c_{2n}He_{2n+2p+1}((J+1/2)/\sigma).
\end{align}\end{subequations}
This can be recast as a Gram-Charlier-like formula
\begin{equation}\label{eq:GGC_KJ}
Q_\text{GC}(J) = \frac{G}{(2\pi)^{1/2}\sigma^2}
  \exp\left(-\frac{(J+1/2)^2}{2\sigma^2}\right)
  \sum_{m=0}^{\infty}d_{2m+1}He_{2m+1}((J+1/2)/\sigma)
\end{equation}
with
\begin{equation}\label{eq:d_vs_c}
d_{2m+1}=\sum_{\substack{n,p\\n+p=m}}\frac{(2\sigma)^{-2p}}{(2p+1)!}c_{2n}.
\end{equation}
The coefficients $d_{2m+1}$ may be obtained by two methods. 
As a general property of the Gram-Charlier expansion (see, e.g., 
Ref.\cite{Pain2020}), the even-order coefficients $c_{2n}$ can be expressed 
versus the cumulants of the $M$-distribution 
\begin{equation}\label{eq:cGC_vs_cumulants}
c_{2n} = \sum_{\substack{a_4,a_6\cdots a_{2n}\\2a_4+3a_6+\cdots+na_{2n}=n}}
  \frac{1}{a_4!}\left(\frac{\kappa_4}{4!\sigma^4}\right)^{a_4}
  \frac{1}{a_6!}\left(\frac{\kappa_6}{6!\sigma^6}\right)^{a_6}\cdots
  \frac{1}{a_{2n}!}\left(\frac{\kappa_{2n}}{(2n)!\sigma^{2n}}\right)^{a_{2n}}.
\end{equation}
The interest of this formula is that the cumulants $\kappa_{2n}$ have 
a fairly simple expression as shown in paper I
\begin{equation}
\kappa_{2k}=\frac{B_{2k}}{2k} \left[\sum_{p=1}^{N}(2j+2-p)^{2k} 
 - \sum_{p=1}^{N}p^{2k} \right]\label{eq:cumul}
\end{equation}
where $B_{2k}$ are the even-order Bernoulli numbers. In addition, when 
several subshells are involved, their cumulants are additive. The second 
method using an alternative expression 
for coefficients $c_{2n}$ is studied in Appendix \ref{sec:psGC_moments}. 

Using the relation (\ref{eq:cGC_vs_cumulants}), one checks that the 
$d_{2m+1}$ coefficient is written as a sum over the indices $(p,a_4,
\dots a_{2n})$ with the constraints $p+n=m$, 
$2a_4+3a_6+\dots n a_{2n}=n$. A thorough analysis of this set of 
indices shows that, after a convenient regrouping of terms, the sum 
is indeed equal to 
\begin{subequations}\begin{align}\label{eq:d_vs_kappa}
d_{2m+1}&=\sum_{\substack{p,a_4\cdots a_{2m}\\p+2a_4+\cdots ma_{2m}=m}}
 \frac{(2\sigma)^{-2p}}{(2p+1)!}
 \frac{1}{a_4!}\left(\frac{\kappa_4}{4!\sigma^4}\right)^{a_4}\cdots
 \frac{1}{a_{2m}!}\left(\frac{\kappa_{2m}}{(2m)!\sigma^{2m}}\right)^{a_{2m}}\\
 &=\frac{1}{\sigma^{2m}}
 \sum_{\substack{p,a_4\cdots a_{2m}\\p+2a_4+\cdots ma_{2m}=m}}
 \frac{2^{-2p}}{(2p+1)!}
 \frac{1}{a_4!}\left(\frac{\kappa_4}{4!}\right)^{a_4}\cdots
 \frac{1}{a_{2m}!}\left(\frac{\kappa_{2m}}{(2m)!}\right)^{a_{2m}}.
\end{align}\end{subequations}
In the above sum, the set $(p,a_4,\cdots a_{2m})$ generates all the partitions 
of the number $m$, as expressed by the condition $p+2a_4+\cdots+ma_{2m}=m$.

The first coefficients $d_{2m+1}$ are
\begin{subequations}\begin{align}
 d_1 &=1\\
 d_3 &= \frac{1}{24\sigma^2}\\\label{eq:d5_PsGC}
 d_5 &= \frac{1}{1920\sigma^4}+\frac{\kappa_4}{24\sigma^4}\\
 d_7 &= \frac{1}{322560\sigma^6}+\frac{\kappa_4}{576\sigma^6}
 +\frac{\kappa_6}{720\sigma^6} 
\end{align}\label{eq:list_coeff_PsGC}\end{subequations}
which can be easily evaluated using the expression of the cumulants for the 
distribution $P(M)$ given by Eq.~(\ref{eq:cumul}). One easily checks that 
accounting only for the first term $d_1$ in the series provides the 
above-mentioned Bethe approximation \cite{Bethe1936} for the $J$ distribution
\begin{equation}\label{eq:Wigner}
 Q_{\text{Bethe}}(J) = \frac{G}{\sigma^3\sqrt{2\pi}}(J+1/2)e^{-\frac{(J+1/2)^2}{2\sigma^2}}.
\end{equation}
The expression (\ref{eq:d_vs_kappa}) for the $d_{2m+1}$ coefficients as 
a function of cumulants is formally simple, however it is also useful to 
get an alternative form as a function of the moments 
$\mu_k=\sum_M M^kP(M)/\sum_M P(M)$. The relevant expression is derived in 
the Appendix \ref{sec:psGC_moments}.

\subsection{Accuracy criterion}
As in paper I, one may define various convergence criteria for the 
obtained Gram-Charlier-like formula. Writing $Q_\text{GC}(J;k)$ 
for the series (\ref{eq:GGC_KJ}) truncated at index $k$ (with $k$ odd), 
the global absolute error is
\begin{equation}\label{eq:abserr}
\Delta_\text{abs}(k)= \left[\sum_{J=J_\text{min}}^{J_\text{max}}
 \Big(Q_\text{GC}(J;k)-Q(J)\Big)^2/(J_\text{max}-J_\text{min}+1)
 \right]^{1/2}
\end{equation}
where $J_\text{min}$ is 0 or 1/2, and the global relative error 
\begin{equation}\label{eq:relerr}
\Delta_\text{rel}(k)= \left[\sum_{J=J_\text{min}}^{J_\text{max}}
 \Big(Q_\text{GC}(J;k)/Q(J)-1\Big)^2/(J_\text{max}-J_\text{min}+1)
 \right]^{1/2}.
\end{equation} 
However, while in paper I the $P(M)$ distribution was always positive 
on its definition range, the number $Q(J)$ may vanish for a series of 
$J$ values. In the relative error computation (\ref{eq:relerr}), these 
values will not be included. Hence the absolute value (\ref{eq:abserr}) 
is here a better accuracy criterion.

\section{Various tests of the Gram-Charlier-like distribution}
\subsection{Single-subshell configuration}
\begin{figure}[htbp]
\centering
\subfigure[Exact and Gram-Charlier-like approximation for the total angular 
momentum distribution $Q(J)$.]%
{\label{fig:QJ_GC_j5n2}%
\includegraphics[scale=0.3,angle=\anglefig]{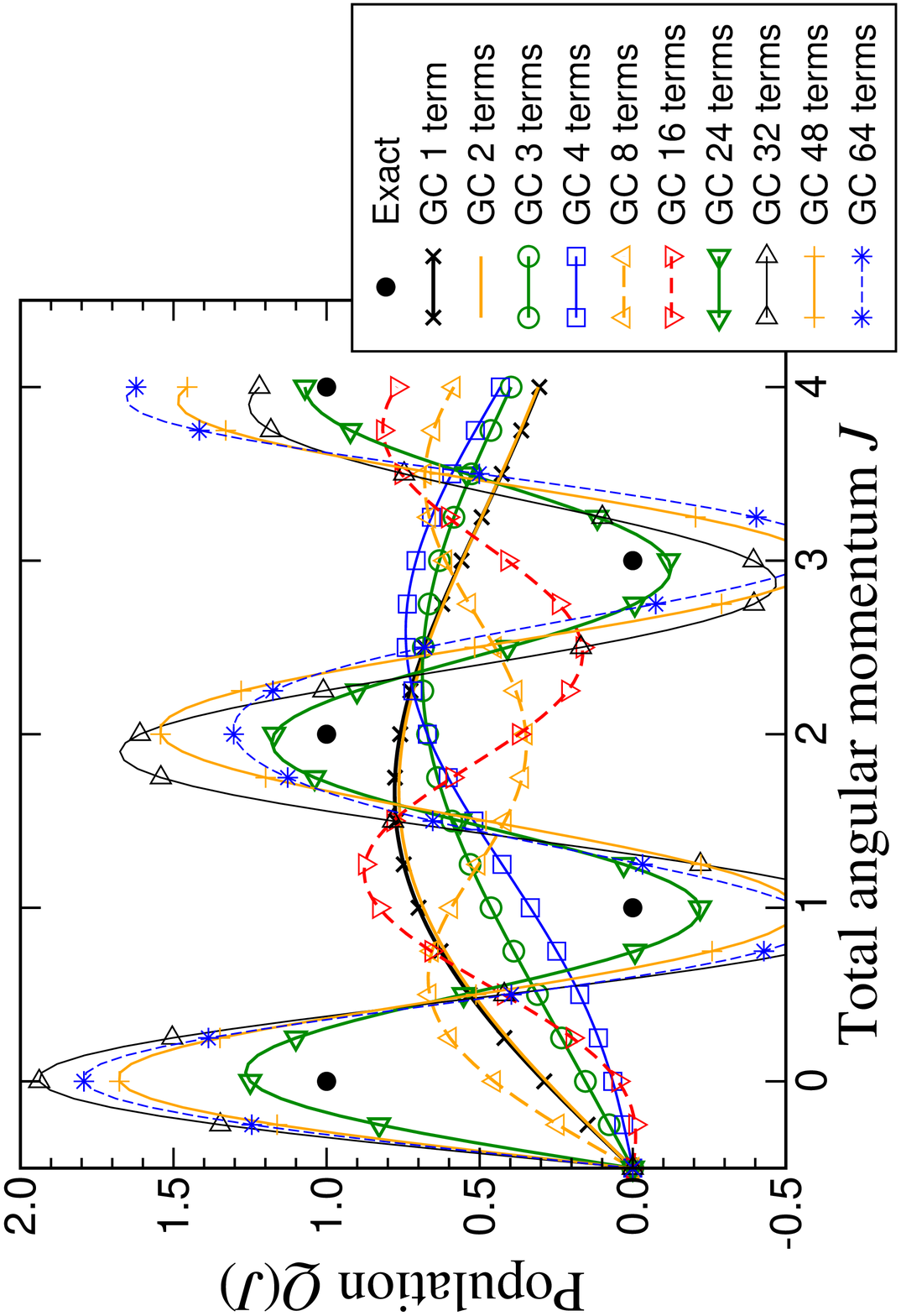}%
}
\hspace{0.1cm}
\subfigure[Error done using the Gram-Charlier-like approximation for the 
total angular momentum distribution $Q(J)$. The abscissa is the 
number of terms kept in the expansion (\ref{eq:GCQJ_sum}).
The contribution of each term to the sum, normalized to the first term, 
is also shown.
]{\label{fig:psGC_j5n2_err}%
\includegraphics[scale=0.3,angle=\anglefig]{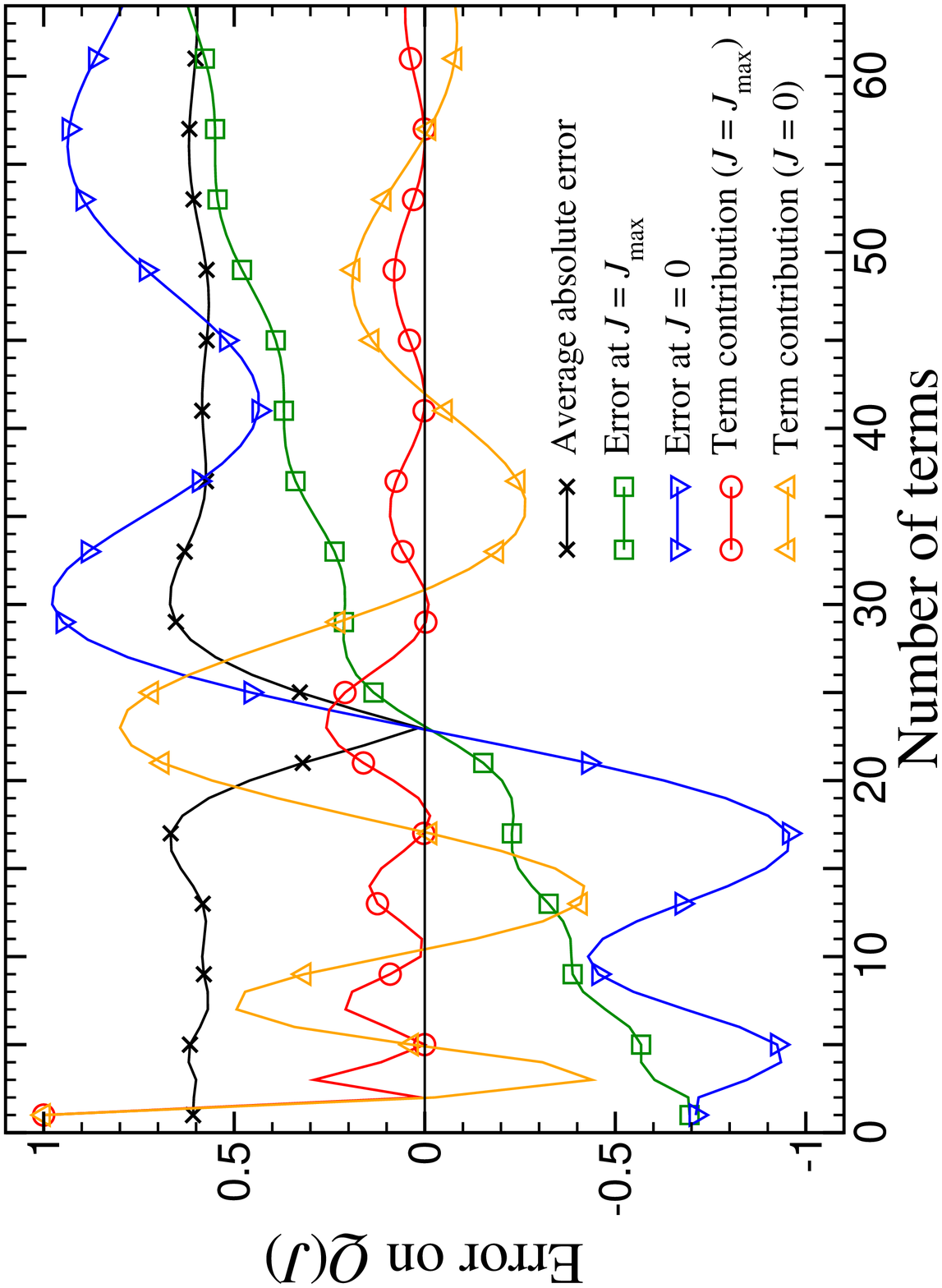}%
}%
\caption{Analysis of the total angular momentum distribution $Q(J)$ in the 
relativistic configuration with one subshell $j_1=5/2, N_1=2$.
}\label{fig:psGC_j5n2}
\end{figure}
Let us first consider a configuration with a small angular momentum 
and number of electrons. On Fig. \ref{fig:QJ_GC_j5n2} we have plotted 
the exact and Gram-Charlier-like approximation for the $J$ distribution 
for two $j=5/2$ electrons. One has then $Q(J)=1$ for $J=0,2,4$ and 
$Q(J)=0$ for $J=1,3$. It is not expected that a Gram-Charlier-like 
expansion provides a reasonable approximation of such a distribution. 
Indeed, we state that a series with few terms provides a very poor 
approximation of $Q(J)$. Quite unexpectedly, we note that including 
about 24 terms in the expansion, the Gram-Charlier-like formula gives 
a reasonable approximation of $Q(J)$. Adding more terms, one notes 
that the quality of the agreement deteriorates gradually. 
In order to analyze the series behavior, we have plotted on 
Fig.~\ref{fig:psGC_j5n2_err} the average absolute error 
(\ref{eq:abserr}) as a function of the number of terms kept in series 
(\ref{eq:GGC_KJ}), e.g., 3 terms corresponds to contributions up to 
$d_5$, and $m$ terms to contributions up to $d_{2m-1}$. 
The average relative error is not plotted here since two $Q(J)$ vanish. 
The absolute error is also shown in the $J=0$ and $J=4$ cases.
We note that the average absolute error is usually about 0.6. This 
quite large value is not surprising since for configurations with few 
electrons the present statistical treatment is not expected to be 
efficient. A more unexpected result is that the error for both 
$J=0$ and $J=4$ cases is small when 23 terms are accounted for. 
Some more information is brought by the analysis of the relative 
contribution of each term, i.e. the ratio 
$\rho(m)=d_{2m-1}He_{2m-1}((J+1/2)/\sigma)/d_1He_1((J+1/2)/\sigma)$
which is also plotted on Fig.~\ref{fig:psGC_j5n2_err} for $J=0$ and 
$J=4$. It appears that this ratio is large for 23 terms ($2m+1=45$), 
which corresponds to the optimum index in the sum. Above this local 
maximum the term contribution decreases but slowly, as a numerical 
analysis shows that the series is not convergent.

The case of a configuration with one subshell and more electrons is 
illustrated by two examples in Appendix \ref{sec:psGC_j15n7-8}. It 
is shown that an acceptable representation of the distribution $Q(J)$ 
is obtained by a Gram-Charlier-like expansion with few terms.

\subsection{Configuration with a small and a large angular momentum}
\begin{figure}[htbp]
\centering
\subfigure[Exact and Gram-Charlier-like approximation for the total 
angular momentum distribution $Q(J)$.]%
{\label{fig:QJ_GC_j5n2j37}%
\includegraphics[scale=0.3,angle=\anglefig]{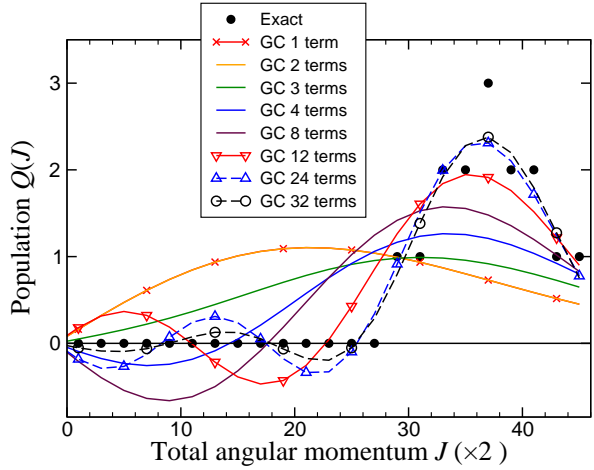}%
}
\hspace{0.1cm}
\subfigure[Error done using the Gram-Charlier-like approximation for the 
total angular momentum distribution $Q(J)$. The abscissa is the 
number of terms kept in the expansion (\ref{eq:GCQJ_sum}).
]{\label{fig:psGC_j5n2j37_err}%
\includegraphics[scale=0.3,angle=\anglefig]{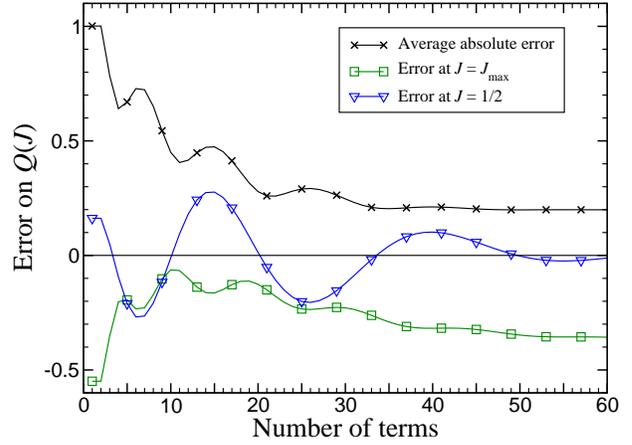}%
}
\caption{Analysis of the total angular momentum distribution $Q(J)$ 
in the relativistic configuration with 2 subshells $j_1=5/2, N_1=2$, 
$j_2=37/2, N_2=1$.}\label{fig:psGC_j5n2j37}
\end{figure}
As a second example, let us consider the $J$ distribution in the configuration 
$j_1=5/2, N_1=2, j_2=37/2, N_2=1$ which is analogous to the case considered 
in Ref.~\cite{Gilleron2009}. It has been shown in paper I that the $M$ distribution 
exhibits a wide plateau for $-29/2\le M\le29/2$, which can be approximated 
by a Gram-Charlier expansion provided a large number of terms is kept. 
Accordingly the $J$ distribution is characterized by a rather sharp peak 
close to $J=37/2$ while all other $Q(J)$ values outside this region cancel. 
Such a stepwise function is certainly very difficult to reproduce with a 
Gram-Charlier-like expansion. Indeed this can be observed on 
Fig.~\ref{fig:QJ_GC_j5n2j37} where we have plotted various Gram-Charlier-like 
approximations versus the exact $Q(J)$ values. It appears that 
the Gram-Charlier-like expansions with one and two terms are 
indistinguishable at the drawing accuracy. Looking at the definitions 
(\ref{eq:list_coeff_PsGC}), one may estimate the first-two-terms ratio
\begin{equation}\label{eq:r2nd1st_psGC}
 \rho(2)=d_3He_3((J+1/2)/\sigma)/d_1He_1((J+1/2)/\sigma)\simeq 
 (J+1/2)^2/(24\sigma^4),
\end{equation}
and for instance at the peak $J=37/2$, using $\sigma^2=1499/12$ obtained from 
Eq. (\ref{eq:sigma2_PM}) this ratio is $\rho\simeq9.6\times10^{-4}$. This 
property holds indeed whatever the configuration analyzed. Looking at the 
various truncated expansions on Fig.~\ref{fig:QJ_GC_j5n2j37}, it appears that 
the three-term expansion including the excess kurtosis $\kappa_4$ is here a 
poor approximation of the $Q(J)$ 
distribution. Including several tens of terms in the expansion, it turns 
out that the series (\ref{eq:GGC_KJ}) ``converges'' toward the correct 
value. However, this appears to be a ``best approximation'' rather than 
a convergence, since adding several hundreds of terms in the series we noted 
that the absolute error levels off at $\simeq0.2$ and does not tend to 0.

The absolute error as defined by (\ref{eq:abserr}) is plotted in 
Fig.~\ref{fig:psGC_j5n2j37_err}, as well as the error done on 
$J_\text{min}=1/2$ and $J_\text{max}=45/2$. As mentioned above, the 
approximations with 1 or 2 terms are almost identical. One notes that  
including high-order terms improves the validity of the Gram-Charlier-like 
expansion. However, the convergence is slow, with a rather limited 
accuracy obtained even when 30 terms are accounted for. This statement 
is in agreement with what was mentioned in paper I for the $P(M)$ 
distribution in the same configuration.

\subsection{Several half-filled subshells}\label{sec:5half-filled}
\begin{figure}[htbp]
\centering
\subfigure[$J$ distribution: exact and Gram-Charlier-like approximation. 
The Gram-Charlier-like expansion includes various terms in 
the sum (\ref{eq:GCQJ_sum}). The 1-term and 2-term approximations are 
indistinguishable at the drawing accuracy, as well as the various 
approximations with more than 2 terms.  
]{\label{fig:QJ_GC_j1j3n2j5n3j7n4j9n5}%
\includegraphics[scale=0.3,angle=\anglefig]%
{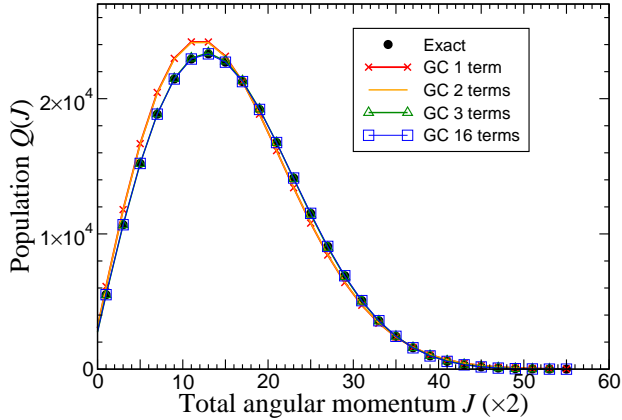}%
}
\hspace{0.1cm}
\subfigure[Error done using the Gram-Charlier-like approximation. 
The average absolute and relative errors are defined in main text. 
The error at $J_0=J_\text{max}$ or $J_0=1/2$ is the value of the 
unsigned difference $|Q_\text{GC}(J_0;k_\text{max})-Q(J_0)|$ plotted 
as a function of the number of terms $(k_\text{max}+1)/2$, where 
$k_\text{max}$ is the truncation index in sum (\ref{eq:GCQJ_sum}).
]{\label{fig:error_GC_j1j3n2j5n3j7n4j9n5}%
\includegraphics[scale=0.3,angle=\anglefig]{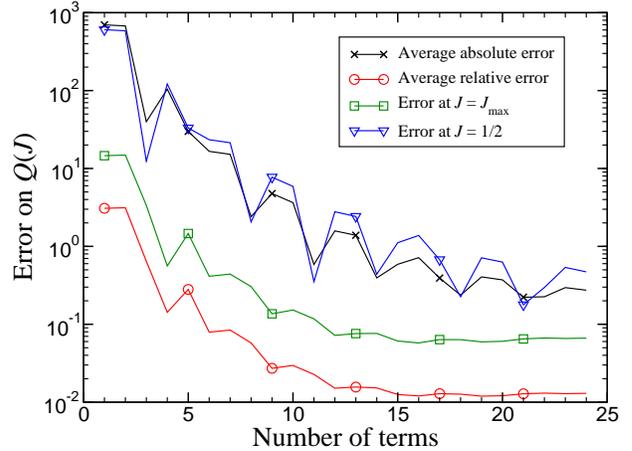}%
}
\caption{Exact and Gram-Charlier-like approximation for the total angular 
momentum $Q(J)$ in the relativistic configuration with 5 half-filled 
subshells $j=i-1/2$, $N_i=j_i+1/2$, $i=1$--$5$.
\label{fig:psGC_j1j3n2j5n3j7n4j9n5}}
\end{figure}
From the central-limit theorem and from paper I analysis we expect 
that the present Gram-Charlier-like approximation will perform better 
when subshells contain a larger number of electrons. To this respect, 
we have plotted in Fig.~\ref{fig:QJ_GC_j1j3n2j5n3j7n4j9n5} the exact 
values of the populations $Q(J)$ and the approximations obtained with 
expansion (\ref{eq:GCQJ_sum}) with various numbers of terms. Only the 
positive part $J>0$ is represented, but of course exact and approximate 
distribution laws verify the rule $Q(-J-1)=-Q(J)$. One observes here 
that the Gram-Charlier-like approximation with 3 terms, including the 
excess kurtosis as shown by Eq. (\ref{eq:d5_PsGC}), is indeed a 
quite good approximation of the distribution $Q(J)$. As mentioned above 
the one- and two-term forms of the expansion (\ref{eq:GCQJ_sum}) are 
almost equal since the ratio (\ref{eq:r2nd1st_psGC}) is again small: 
$1/24\sigma^2=1/1010$. Fig.~\ref{fig:error_GC_j1j3n2j5n3j7n4j9n5} is a 
plot of various expressions of the error as a function of the number of 
terms retained in the expansion (\ref{eq:GCQJ_sum}). This second part 
of Fig.~\ref{fig:psGC_j1j3n2j5n3j7n4j9n5} shows that, while including 
3 terms in the expansion is an acceptable approximation, much better 
results may be obtained with more terms, the optimum being reached with 
about 12 terms. Though a mathematical analysis of the convergence of 
the series (\ref{eq:GCQJ_sum}) would be outside the scope of this work, 
we may conclude that such expansion is of asymptotic nature. 

\subsection{Configuration with a large number of subshells}
\begin{figure}[htbp]
\centering
\subfigure[$J$ distribution: exact and Gram-Charlier-like approximation 
]{\label{fig:QJGC_j1j3-j19}%
\includegraphics[scale=0.3,angle=\anglefig]{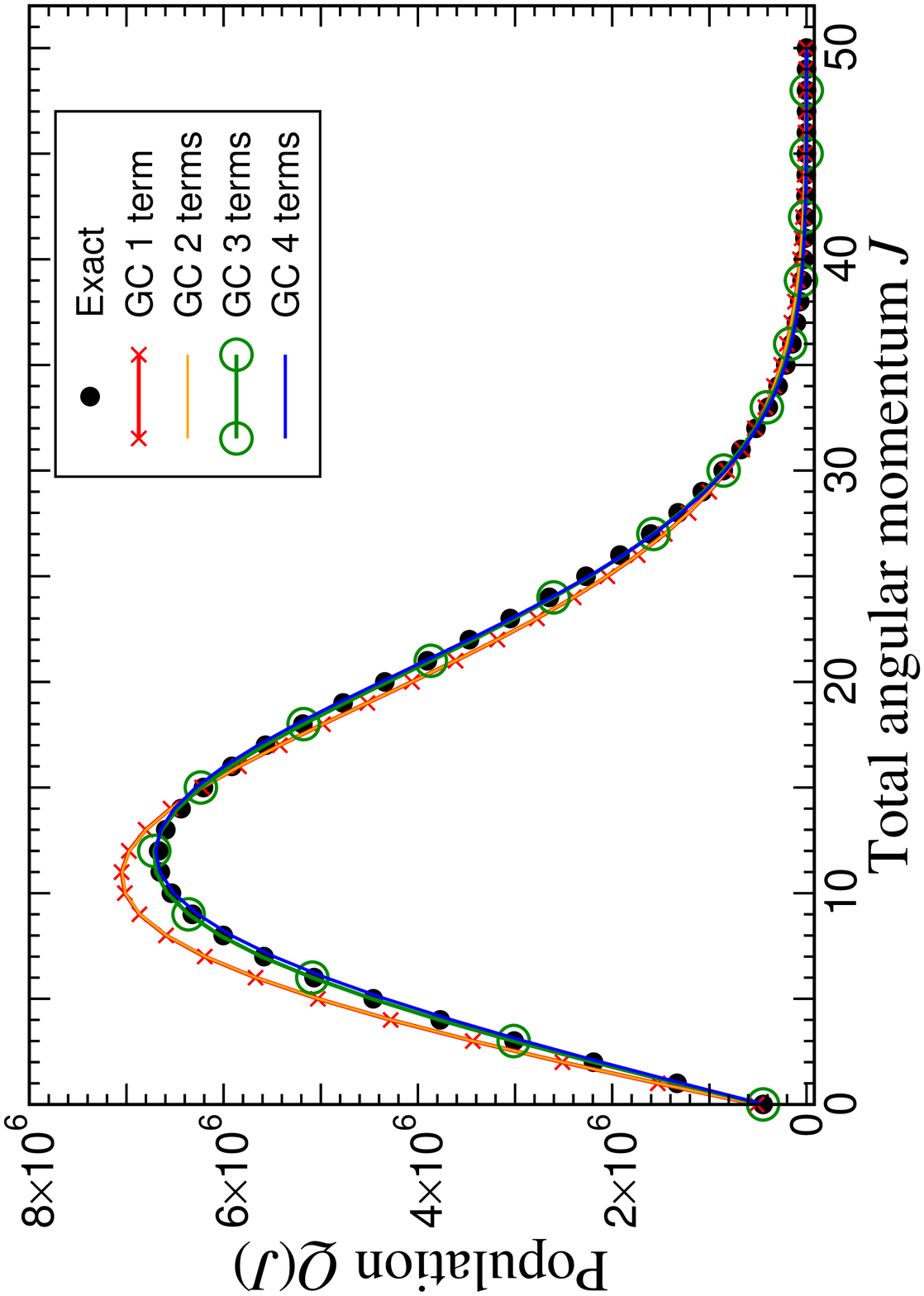}%
}
\hspace{0.1cm}
\subfigure[Difference between the populations at various orders and 
exact value. For the first two cases, the difference is divided by 
10.]{\label{fig:DeltaQJ_j1j3-j19}%
\includegraphics[scale=0.3,angle=\anglefig]{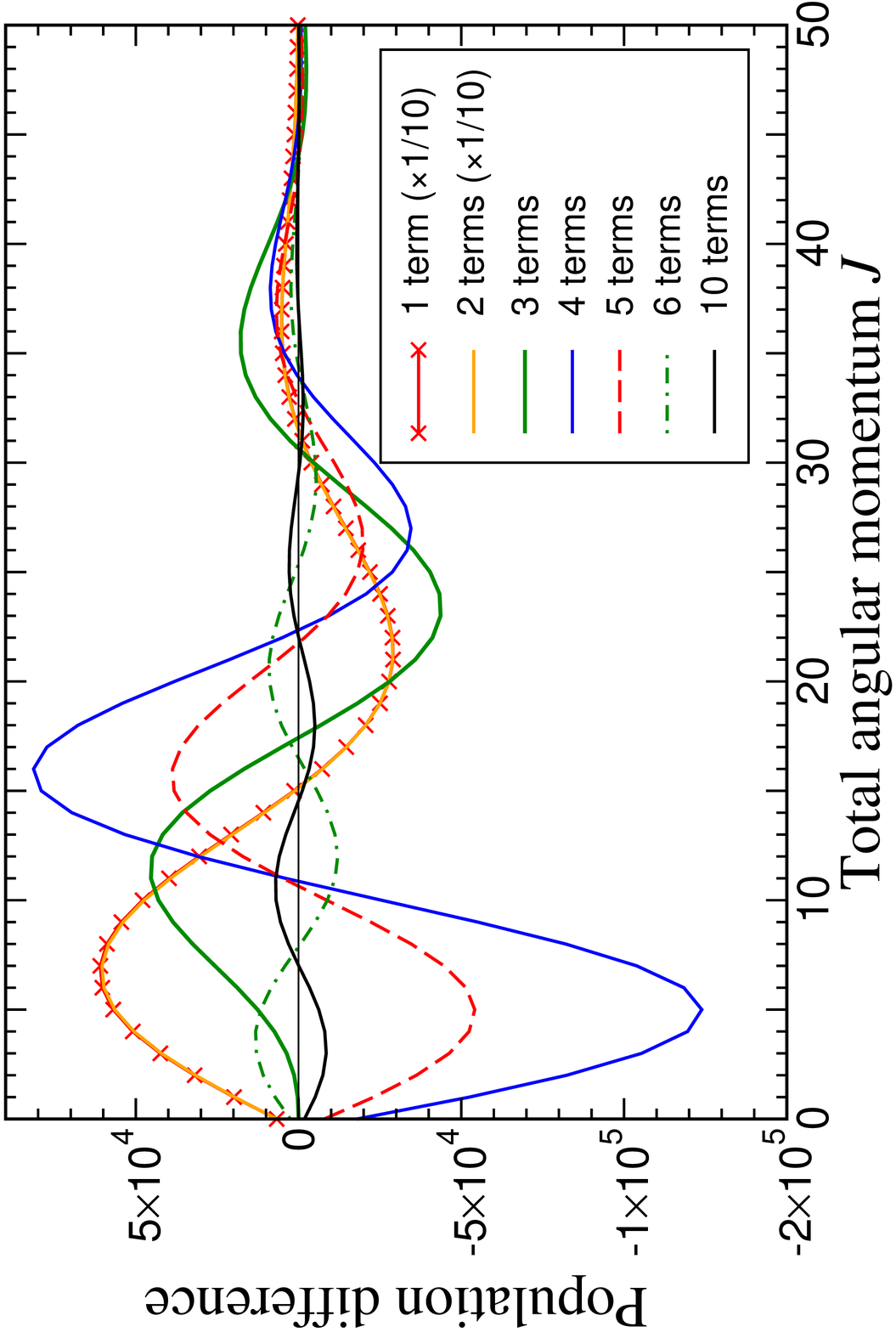}%
}\\
\subfigure[Error done on $Q(J)$ as a function of the number of terms kept
in the expansion (\ref{eq:GGC_KJ}). 
See Fig.~\ref{fig:error_GC_j1j3n2j5n3j7n4j9n5} caption for details.
]{\label{fig:error_GC_j1j3-j19}%
\includegraphics[scale=0.3,angle=\anglefig]{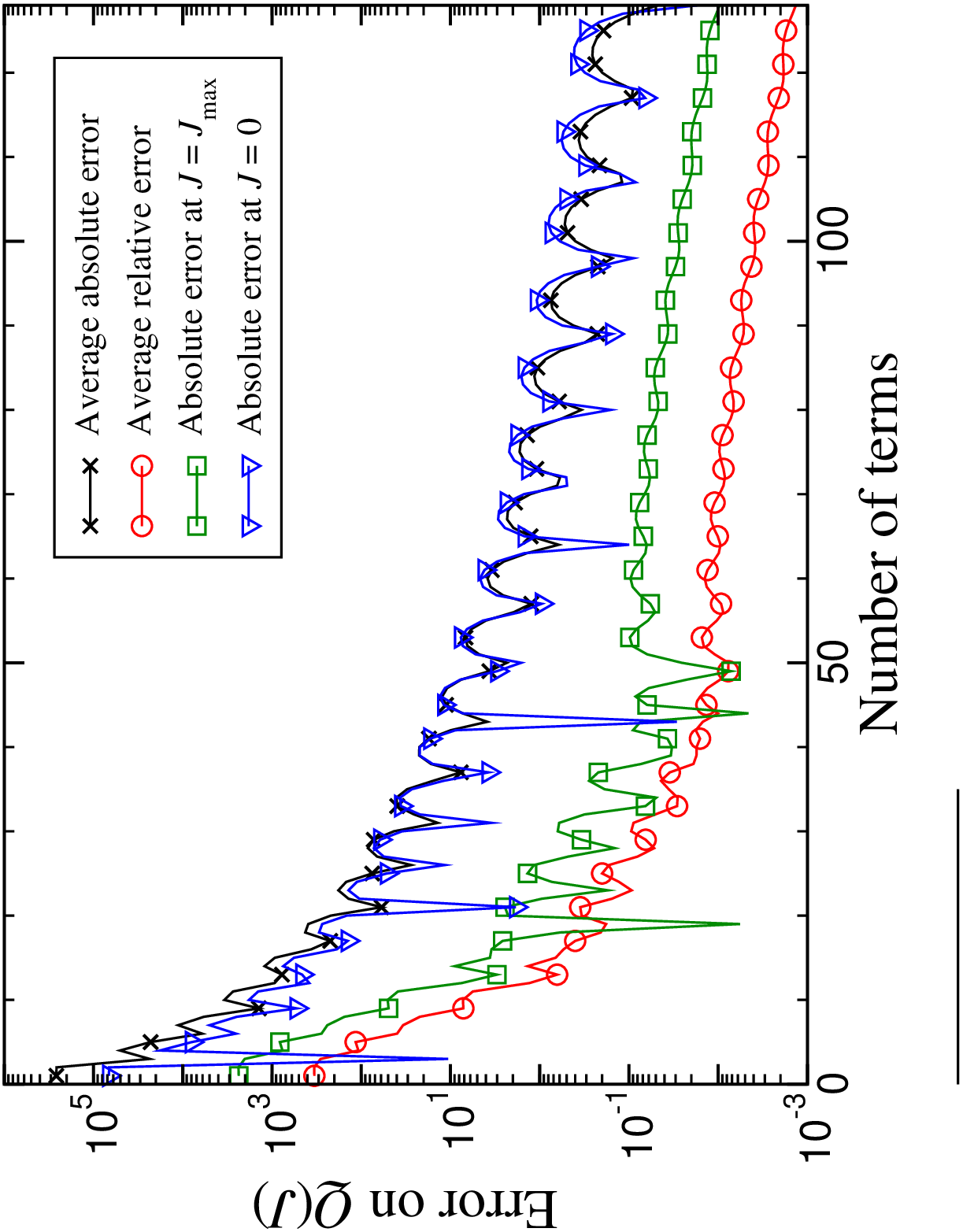}}%
\caption{Exact and Gram-Charlier-like approximation for the total angular 
momentum distribution $Q(J)$ in the relativistic configuration with 10 
subshells $j_i=i-1/2, i=1\text{--}10$ all singly populated.
\label{fig:QJ_GC_j1j3-j19}}
\end{figure}

As in paper I, we consider here the distribution $Q(J)$ in the case of a 
configuration with 10 subshells $j=1/2$--$19/2$, all containing a single 
electron. For this 10-electron configuration one has $J_\text{max}=50$, 
the degeneracy is $2^{10}.10!=3.7158912\times10^9$, and the population 
$P(M)$ varies on 8 orders of magnitude --- $Q(J)$ varying on 7 orders of 
magnitude. We have plotted in 
Fig.~\ref{fig:QJ_GC_j1j3-j19} the $Q(J)$ distribution computed exactly and 
the Gram-Charlier-like expansions truncated at various orders. 
One observes once again that the approximations including one or two terms 
are very similar, and that in this case both differ hardly from the exact 
value. Conversely, as was observed for the $M$-distribution 
\cite{Poirier2021}, the expansions including at least 3 terms, i.e., 
the kurtosis contribution or term $d_5$ in Eq.(\ref{eq:d_vs_kappa}), 
provide a fair representation of the angular momentum distribution on the 
whole range.
As seen on Fig.~\ref{fig:error_GC_j1j3-j19}, the 3-term approximation 
brings a significant improvement with respect to the Bethe formula 
(1-term approximation), while at least 6 terms are necessary to get 
a better accuracy. The average error versus the number of terms is 
plotted on Fig.~\ref{fig:error_GC_j1j3-j19} which shows the accuracy 
slowly improves with the number of terms. 

In order to confirm the non-convergent character of the 
Gram-Charlier expansion, we performed several Gram-Charlier-expansion 
studies on configurations with $N$ singly-populated subshells $j_i=i-1/2, 
i\text{ from }1\text{ to }N, N_i=1$ for each subshell. For $N=5$, we 
observed that the smallest relative error (\ref{eq:relerr}) was reached for 
92 terms in the series, with an error $\Delta_\text{rel}=3\times10^{-5}$. 
For $N=6$ (resp.~8), the minimum occurs for 148 (resp.~334) terms and 
reaches $\Delta_\text{rel}=10^{-5}$ (resp.~$3\times10^{-7}$). Beyond these 
optimal values, the error tends to increase. We therefore note that the 
larger the number of electrons, the later the onset of the divergence, and 
the better the quality of the Gram-Charlier expansion. A similar behavior 
occurs for the Stirling expansion for the factorial. Such series are called 
asymptotic and are studied in various treatises \cite{Erdelyi1956}.

\section{Application to the computation of moments at any order}
In this section, we consider the case of a single-subshell configuration 
$j^N$. The present conclusions may be generalized to the multiple-subshell 
case by taking benefit of the additivity of cumulants --- the value 
$\kappa_{2k}$ for a single subshell is to be replaced by the sum of the 
contributions of individual subshells. 
The formula (\ref{eq:GGC_KJ}) is useful to get the odd-order moments, 
i.e, the average of $(2J+1)^{2n+1}$ on the distribution $Q(J)$. The first 
odd-order moment, that will be used as the normalization factor, is simply 
equal to the total degeneracy of the configuration
\begin{equation}
  \sum_{J=J_\text{min}}^{J_\text{max}}(2J+1)Q(J) = \binom{2j+1}{N} = G.
\end{equation}
In order to evaluate the higher odd-order moments, one uses the classical 
inversion property of the Hermite polynomials --- easily derived using the 
Rodrigues formula (\ref{eq:Hermite}) and repeated integration by parts ---
\begin{equation}\label{eq:Hermite_inv}
X^{k}= k!\sum_{m=0}^{\lfloor k/2\rfloor}\frac{He_{k-2m}(X)}{2^m m!(k-2m)!},
\end{equation}
where $\lfloor x\rfloor$ is the integer part of $x$, and the orthogonality 
property \cite{Abramowitz1972}
\begin{equation}
\int_{-\infty}^{+\infty}\: dX\exp(-X^2/2)He_m(X)He_n(X) = 
 \delta_{nm}\sqrt{2\pi}n!
\end{equation}
which allows us to get directly the average values $\left<He_{2n+1}(Q+1/2)
\right>$ on the Gram-Charlier-like distribution (\ref{eq:GGC_KJ}). 
Replacing the discrete sum over $J$ by an integral $\int_{-1/2}^\infty$ where 
the symmetry property $Q_\text{GC}(-J-1)=-Q_\text{GC}(J)$ (meaning that 
the function $Q_\text{GC}(J)$ is odd versus the origin $J=-1/2$) is used, one 
gets at once
\begin{subequations}\begin{align}
 \label{eq:intQ}\sum_J He_{2n+1}((J+1/2)/\sigma)Q(J)& \simeq
 \int_{-1/2}^{+\infty} dJ Q_\text{GC}(J)He_{2n+1}((J+1/2)/\sigma)\\ 
  &=\frac12\int_{-\infty}^{+\infty} dJ Q_\text{GC}(J)He_{2n+1}((J+1/2)/\sigma)\\
  &= \frac12\frac{G}{\sigma} (2n+1)! d_{2n+1}
\end{align}\end{subequations}
where the $d_{2n+1}$ are known as a function of the cumulants using 
Eq. (\ref{eq:d_vs_kappa}). In some cases one can even show that the 
equality (\ref{eq:intQ}) is indeed not approximate but exact.
Using the inversion formula (\ref{eq:Hermite_inv}), one has also
\begin{subequations}\begin{align}
\sum_J \left((J+1/2)/\sigma)\right)^{2n+1}Q(J)& \simeq
 \int_{-1/2}^{+\infty} dJ Q_\text{GC}(J)((J+1/2)/\sigma)^{2n+1}\\ 
  &= \frac12\frac{G}{\sigma} (2n+1)!
  \sum_{m=0}^n \frac{d_{2m+1}}{2^m m!}.
\end{align}\end{subequations}

More generally, using the formula 
\begin{equation}\int_0^\infty y^{p}e^{-y^2/2}He_{n}(y)dy  = 
\frac{\Gamma(p+1)2^{(n-p-1)/2}\sqrt{\pi}}{\Gamma(1+(p-n)/2)}
\end{equation}
which can be derived using the formula (\ref{eq:Hermite}) and repeated 
integration by parts, one may easily obtain an approximation for the sum 
$\sum_{J\ge0} \left((2J+1)/\sigma)\right)^{n}Q(J)$ including for even $n$. 
Explicitly, one gets 
\begin{equation}\label{eq:mtQpos}
M_n = \sum_{J\ge0} (2J+1)^{n}Q(J) \simeq
 \int_{-1/2}^\infty (2J+1)^{n} Q_\text{GC}(J)dJ = 
2G (2\sigma)^{n-1}n!\sum_{m\ge0}
 \frac{2^{m-(n+1)/2}}{\Gamma\left(\frac{n+1}{2}-m\right)}d_{2m+1}.
\end{equation}
If $n$ is odd, the above sum terminates at $m=(n-1)/2$. Moreover, one 
can prove that the relation (\ref{eq:mtQpos}) is not approximate but exact
at least concerning odd orders for which convergence is not an issue.
The first odd-order moments are
\begin{subequations}\begin{align}
M_1 &= G\\
M_3 &= G(1+12\sigma^2)\\
M_5 &= G(1+40\sigma^2+240\sigma^4+80\kappa_4)\\
M_7 &= G(1+84\sigma^2+1680\sigma^4+6720\sigma^6+
 560(1+12\sigma^2)\kappa_4+448\kappa_6).
\end{align}\end{subequations}
Various examples of $M_n$ values at even and odd order $n$ are presented in 
Table~\ref{tab:moments_Mn}.
\begin{table}[htbp]
\renewcommand{\arraystretch}{1.25}
 \centering
 \begin{tabular}{@{\quad}*{8}{c}}
 \hline\hline
 $n$  &  Exact    &  1 term      &    2 terms     &    3 terms    &    4 terms    &   8 terms    &  16 terms\\
 \hline
  0 &         468 &        483.094 &        482.869 &    468.728    &       466.171 &      468.715 &      468.513 \\
  1 &       11440 &      11440     &      11440     &     11440     &     11440     &    11440     &    11440    \\
  2 &      348584 &     344929     &     345090     &    348456     &    348821     &   348575     &   348585 \\
  3 & 1.226368$\times10^7$ & 1.225224$\times10^7$ & 1.226368$\times10^7$ & 1.226368$\times10^7$ & 1.226368$\times10^7$ & 1.226368$\times10^7$ & 1.226368$\times10^7$ \\
  4 & 4.785009$\times10^8$ & 4.925592$\times10^8$ & 4.932490$\times10^8$ & 4.788311$\times10^8$ & 4.783097$\times10^8$ & 4.785042$\times10^8$ & 4.785008$\times10^8$ \\
  5 & 2.020396$\times10^{10}$ & 2.187025$\times10^{10}$ & 2.191109$\times10^{10}$ & 2.020396$\times10^{10}$ & 2.020396$\times10^{10}$ & 2.020396$\times10^{10}$ & 2.020396$\times10^{10}$ \\
  6 & 9.080727$\times10^{11}$ & 1.055062$\times10^{12}$ & 1.057525$\times10^{12}$ & 9.031082$\times10^{11}$ & 9.086918$\times10^{11}$ & 9.080691$\times10^{11}$ & 9.080727$\times10^{11}$ \\
  7 & 4.293402$\times10^{13}$ & 5.465375$\times10^{13}$ & 5.480684$\times10^{13}$ & 4.200846$\times10^{13}$ & 4.293402$\times10^{13}$ & 4.293402$\times10^{13}$ & 4.293402$\times10^{13}$ \\
 \hline\hline
 \end{tabular}
 \caption{Moments $M_n$ for the distribution $j=15/2$, $N=7$ as a function 
 of the number of terms kept in the series (\ref{eq:mtQpos}). Odd-order 
 values are obtained using a finite sum.}
 \label{tab:moments_Mn}
\end{table}

An interesting consequence of this formula is an approximation for the useful 
quantity $M_0=\sum_{J\ge0} Q(J)$, representing the total number of levels. 
Recently, a calculation of $M_0$ using fractional parentage coefficients was 
proposed for $j^3$ and $j^4$ configurations \cite{Pain2019}. In the former 
case, it was found that $M_0=(4j^2-1)/8$, and in the latter an expression of 
$M_0$ as a polynomial in $j$ plus a triple sum of products of $6j$ 
coefficients was obtained. To our 
knowledge, no simple analytical expression of the total number of levels 
is available for a general configuration, even with a single subshell. 
One has
\begin{subequations}\label{eq:M0_form1}\begin{align}
M_0&=\sum_{J\ge0} Q(J)
\simeq \frac{G}{(2\pi)^{1/2}\sigma}
 \sum_{m\ge0} \frac{2^m\Gamma(1/2)}{\Gamma(1/2-m)}d_{2m+1}\\
 &= \frac{G}{(2\pi)^{1/2}\sigma}
 \left(1+\sum_{m>0}(-1)^m(2m-1)!!d_{2m+1}\right)\\
 &= \frac{G}{(2\pi)^{1/2}\sigma}
 \left[1 -\frac{1}{24\sigma^2} + \left(
 \frac{1}{640\sigma^4}+\frac{\kappa_4}{8\sigma^4} \right)
 -\left(\frac{1}{21504\sigma^6}+\frac{5\kappa_4}{192\sigma^6}
 +\frac{\kappa_6}{48\sigma^6}\right)+\cdots
 \right]
\end{align}\end{subequations}
where $(2m-1)!!=1.3.5\cdots(2m-1)$ is the double factorial. As mentioned 
concerning the series (\ref{eq:GGC_KJ}), we do not claim that such expansion 
is convergent, but it is likely of asymptotic nature. 
An alternative expression for $M_0$ is provided by Eq. (\ref{eq:QJvsPM}). 
One has for even $N$, using paper I results for the Gram-Charlier expansion 
of $P(M)$,
\begin{subequations}\label{eq:M0_form2}\begin{equation}
  M_0=P(0)=\frac{G}{(2\pi)^{1/2}\sigma}\left(1+
   \sum_{n>1}(-1)^n(2n-1)!!c_{2n}\right)\text{\quad(even $N$)}\label{eq:P0_GC}
\end{equation}
where the value $He_{2n}(0)=(-1)^n(2n)!/2^nn!=(-1)^n(2n-1)!!$ has been used. 
The sum of $Q(J)$ for odd $N$ is more complicated and is written as
\begin{equation}
  M_0=P(1/2)=\frac{G}{(2\pi)^{1/2}\sigma}\exp(-1/8\sigma^2)
   \left(1+\sum_{n>1}c_{2n}He_{2n}(1/2\sigma)\right)
   \text{\quad(odd $N$)}.\label{eq:M0_vsPh}
\end{equation}\end{subequations}
The approximations (\ref{eq:M0_form1},\ref{eq:M0_form2}) are compared in 
Tables \ref{tab:approx_M0_C12} and \ref{tab:approx_M0_C34} 
as a function of the number of terms in the 
expansion. The cutoff index $k_\text{max}$ is directly related to the number 
of terms involved in the sums. It corresponds to the maximum order of the 
cumulant $\kappa(k_\text{max})$, for instance if $k_\text{max}=4$, cumulants 
up to $\kappa_4$ are accounted for. If the form (\ref{eq:M0_form1}) is used, 
the sum contains $k_\text{max}/2+1$ terms. If the form (\ref{eq:M0_form2}) 
is used, the sum contains $k_\text{max}/2$ terms. 

These approximations are first tested in Table~\ref{tab:approx_M0_C12} 
for the single-subshell configuration $7/2^3$, for which the exact value 
is $M_0=6$. We note that the error slowly decreases with $k_\text{max}$, 
and that both forms oscillate. A rather good approximation is reached with 
4 terms included in the sum.
As a second example we consider the two-subshell configuration $5/2^2 37/2$, 
for which the populations $Q(J)$ are plotted in Fig.~\ref{fig:QJ_GC_j5n2j37}. 
As seen on Table~\ref{tab:approx_M0_C12}, both approximations are of the 
same quality whatever $k_\text{max}$. Taking only 3 terms --- up to the 
kurtosis $\kappa_4$ --- provides a poor approximation of the sum $\sum_JQ(J)$. 
With one more term the approximation improves, but this behavior is particular 
to this case. As a rule, for configurations with a small number of electrons, 
one notes that the error for both forms oscillates with $k_\text{max}$ and 
tends to decrease, though quite slowly. It appears from this table that 
about 30 terms are needed to get an accuracy well below the percent level.
In Table~\ref{tab:approx_M0_C34}, case 3 is the configuration composed of 
half-filled subshells $j_i=i-1/2, N_i=i$ with $i$ from 1 to 5.
It appears that the second form, involving the Gram-Charlier expression 
for $P(J_\text{min})$, provides a much better approximation for this sum. 
This is probably due to the inaccuracy added by replacing the difference 
$P(J)-P(J+1)$ by a derivative in defining (\ref{eq:M0_form1}). 
One notices that the accuracy Eq.~(\ref{eq:M0_form1}) reaches a plateau at 
$\simeq0.002$ while the error using Eq.~(\ref{eq:M0_form2}) decreases.
One notes that the approximation labeled as $k_\text{max}=4$, which is 
the first term involving the excess kurtosis $\kappa_4$, 
can be significantly improved by adding more terms in the 
series if the second form (\ref{eq:M0_form2}) is used.
The same table presents the results for the 10-subshell configuration 
corresponding to Fig.~\ref{fig:QJ_GC_j1j3-j19} analysis. The conclusions 
obtained in Case-3 configuration apply here too.

\begin{table}[htbp]
 \centering
 \begin{tabular}{@{\extracolsep{4pt}}@{\quad}c@{\ }cccc cccc}
 \hline\hline
  & \multicolumn{4}{c}{Case 1} & \multicolumn{4}{c}{Case 2} \\
\cline{2-5} \cline{6-9}
 $k_\text{max}$ & Form 1 &  error &  Form 2 & error & Form 1 &  error &  Form 2 & error \\
 \hline
  0 &  6.66073 &  1.10(-1) & 6.58713 &  9.79(-2) & 20.3458 &  3.56(-1) & 20.3254 &  3.55(-1) \\
  2 &  6.63606 &  1.06(-1) & 6.58713 &  9.79(-2) & 20.3390 &  3.56(-1) & 20.3254 &  3.55(-1) \\
  4 &  6.21430 &  3.57(-2) & 6.18842 &  3.14(-2) & 17.5044 &  1.67(-1) & 17.5050 &  1.67(-1) \\
  6 &  6.05749 &  9.58(-3) & 6.03639 &  6.07(-3) & 14.9108 & -5.94(-3) & 14.9249 & -5.01(-3) \\
  8 &  6.10829 &  1.80(-2) & 6.07842 &  1.31(-2) & 13.2786 & -1.15(-1) & 13.3013 & -1.13(-1) \\
 10 &  6.22233 &  3.71(-2) & 6.18026 &  3.00(-2) & 12.6002 & -1.60(-1) & 12.6255 & -1.58(-1) \\
 12 &  6.30396 &  5.07(-2) & 6.25445 &  4.24(-2) & 12.6392 & -1.57(-1) & 12.6615 & -1.56(-1) \\
 16 &  6.27282 &  4.55(-2) & 6.23351 &  3.89(-2) & 13.8113 & -7.92(-2) & 13.8172 & -7.89(-2) \\
 24 &  6.01031 &  1.72(-3) & 6.01452 &  2.42(-3) & 15.9746 &  6.50(-2) & 15.9508 &  6.34(-2) \\
 32 &  6.13742 &  2.29(-2) & 6.10293 &  1.72(-2) & 15.8600 &  5.73(-2) & 15.8438 &  5.63(-2) \\
 64 &  5.86148 & -2.31(-2) & 5.94881 & -8.53(-3) & 14.9298 & -4.68(-3) & 14.9297 & -4.69(-3) \\
128 &  5.98820 & -1.97(-3) & 5.97753 & -3.74(-3) & 15.0072 &  4.77(-4) & 15.0065 &  4.33(-4) \\
 \hline\hline
 \end{tabular}
\caption{Values of the moment $M_0=\sum_{J\ge0}Q(J)$ using formulas 
(\ref{eq:M0_form1},\ref{eq:M0_form2}) as Form 1, Form 2 respectively. 
The error is the relative variation $M_0(\text{approx})/M_0(\text{exact})
-1$. The value $k_\text{max}$ for each row corresponds to the cutoff 
index (see text for details). 
Case 1 is the single-subshell configuration $7/2^3$, for which $M_0=6$. 
Case 2 is the configuration with 2 subshells 
$5/2^2 37/2^1$, for which the exact value is $M_0=15$. 
The notation $a(-b)$ stands for $a\times10^{-b}$.
}
\label{tab:approx_M0_C12}
\end{table}
\begin{table}[htbp]
 \centering
 \begin{tabular}{@{\extracolsep{4pt}}@{\quad}c@{\ }*{8}{c}}
 \hline\hline
  & \multicolumn{4}{c}{Case 3} 
  & \multicolumn{4}{c}{Case 4}\\
\cline{2-5} \cline{6-9}
 $k_\text{max}$ & 
  Form 1 &  error &  Form 2 & error & 
  Form 1 &  error &  Form 2 & error\\
 \hline
  0 & 260354.3 &  2.36(-2) & 259582.2 &  2.06(-2) &
    131285863 &  2.80(-2) & 131285863 &  2.80(-2) \\
  2 & 260096.6 &  2.26(-2) & 259582.2 &  2.06(-2) &
    131242959 &  2.77(-2) & 131285863 &  2.80(-2) \\
  4 & 254912.7 &  2.19(-3) & 254474.8 &  4.71(-4) &
    127833224 &  9.86(-4) & 127876115 &  1.32(-3) \\
  6 & 254232.3 & -4.82(-4) & 253783.3 & -2.25(-3) &
    127175754 & -4.16(-3) & 127213074 & -3.87(-3) \\
  8 & 254664.9 &  1.22(-3) & 254199.8 & -6.10(-4) &
    127462387 & -1.92(-3) & 127498193 & -1.64(-3) \\
 10 &  254892.6 &  2.11(-3) & 254423.7 &  2.70(-4) &
    127705777 & -1.19(-5) & 127742424 &  2.75(-4) \\
 12 & 254885.8 &  2.09(-3) & 254419.6 &  2.54(-4) &
    127754702 &  3.71(-4) & 127792226 &  6.65(-4) \\
 16 & 254801.5 &  1.76(-3) & 254338.3 & -6.58(-5) &
    127665972 & -3.24(-4) & 127703495 & -2.98(-5) \\
 24 & 254823.8 &  1.84(-3) & 254359.2 &  1.66(-5) &
    127673721 & -2.63(-4) & 127711002 &  2.90(-5) \\
 32 & 254818.4 &  1.82(-3) & 254354.1 & -3.71(-6) &
    127669135 & -2.99(-4) & 127706465 & -6.56(-6) \\
 64 & 254819.1 &  1.82(-3) & 254354.8 & -9.53(-7) &
    127670125 & -2.91(-4) & 127707441 &  1.08(-6) \\
 \hline\hline
 \end{tabular}
\caption{Values of the moment $M_0=\sum_{J\ge0}Q(J)$ using formulas 
(\ref{eq:M0_form1},\ref{eq:M0_form2}). 
Case 3 is the configuration with 5 half-filled subshells $1/2^13/2^2
5/2^37/2^49/2^5$, the exact value of the moment is $M_0=254355$. 
Case 4 is the configuration with 10 subshells $j_i=1/2,3/2\dots19/2$ 
all singly populated, with exact value $M_0=127707302$.
See Table \ref{tab:approx_M0_C12} for details.
}
\label{tab:approx_M0_C34}
\end{table}

\section{Number of lines in a transition array}
Among useful applications of the counting of levels with a given total angular 
momentum $J$, one finds for instance the numbering of Auger amplitudes, for 
which Kyni\.ene \textit{et al} obtained a fair approximation \cite{Kyniene2002}, 
or the determination of the number of radiative transitions between two relativistic 
configurations. We deal here with the case of dipolar transitions (E1 or M1) 
between two relativistic configurations $A$ and $B$. The number of such lines 
is (see, e.g. Ref.\cite{Bauche1987})
\begin{equation}\label{eq:ntransDE}
N_{AB}=\sum_J{}'\;\Big(Q_A(J)Q_B(J)+Q_A(J)Q_B(J+1)+Q_A(J+1)Q_B(J)\Big)
\end{equation}
where $Q_A(J)$ (resp.~$Q_B(J)$) is the number of levels with total angular 
momentum $J$ in configuration $A$ (resp.~$B$). The prime in the sum reflects 
the fact that it is restricted to non-negative $J$ values, and that one must 
eliminate the 0--0 transitions if the total number of electrons is even, 
therefore one has to subtract $Q_A(0)Q_B(0)$ from the sum.
\begin{table}[htbp]
 \centering
 \begin{tabular}{@{\extracolsep{4pt}}@{\quad}c@{\ }*{10}{c}}
 \hline\hline
  & \multicolumn{2}{c}{Case 1} & \multicolumn{2}{c}{Case 2} & \multicolumn{2}{c}{Case 3} 
  & \multicolumn{2}{c}{Case 4} & \multicolumn{2}{c}{Case 5}\\
\cline{2-3} \cline{4-5} \cline{6-7} \cline{8-9} \cline{10-11}
 $k_\text{max}$ & Count & error & Count & error & Count &  error 
   & Count &  error & Count &  error \\
 \hline
Exact  &       106 &           &        79 &           &       189 &           &     10983 &           &  3925292 &          \\
   1   &  121.2746 &  1.44(-1) &  46.01977 & -4.18(-1) &  222.9916 &  1.80(-1) &  12118.94 &  1.03(-1) &  4260164 &  8.53(-2)\\
   3   &  120.6034 &  1.37(-1) &  45.97171 & -4.18(-1) &  219.6235 &  1.62(-1) &  12060.55 &  9.81(-2) &  4237309 &  7.95(-2)\\
   5   &  105.8427 & -1.48(-3) &  33.53130 & -5.76(-1) &  191.7090 &  1.43(-2) &  11064.90 &  7.46(-3) &  3947621 &  5.69(-3)\\
   7   &  103.8143 & -2.06(-2) &  41.57908 & -4.74(-1) &  186.4600 & -1.34(-2) &  10945.10 & -3.45(-3) &  3911716 & -3.46(-3)\\
   9   &  103.9086 & -1.97(-2) &  58.53235 & -2.59(-1) &  186.4040 & -1.37(-2) &  10969.21 & -1.26(-3) &  3920213 & -1.29(-3)\\
  15   &  104.4607 & -1.45(-2) &  62.29300 & -2.12(-1) &  187.5295 & -7.78(-3) &  10986.25 &  2.96(-4) &  3925469 &  4.51(-5)\\
  23   &  104.6794 & -1.24(-2) &  69.30309 & -1.23(-1) &  187.5285 & -7.79(-3) &  10985.39 &  2.18(-4) &  3925279 & -3.41(-6)\\
  31   &  104.4558 & -1.45(-2) &  75.71697 & -4.16(-2) &  187.5797 & -7.52(-3) &  10985.35 &  2.14(-4) &  3925281 & -2.72(-6)\\
  47   &  104.3896 & -1.51(-2) &  77.91521 & -1.37(-2) &  187.7243 & -6.75(-3) &  10985.36 &  2.15(-4) &  3925283 & -2.37(-6)\\
  63   &  104.3722 & -1.53(-2) &  76.75403 & -2.84(-2) &  188.2175 & -4.14(-3) &  10985.39 &  2.18(-4) &  3925284 & -1.97(-6)\\
 127   &  104.7995 & -1.13(-2) &  77.78638 & -1.54(-2) &  191.8610 &  1.51(-2) &  10985.43 &  2.21(-4) &  3925288 & -9.46(-7)\\
 \hline\hline
 \end{tabular}
\caption{Number of lines obtained with the Gram-Charlier-like expansion as 
a function of the cutoff index $k_\text{max}$ for various transition arrays. 
Case 1: $9/2^3\rightarrow9/2^2 11/2$, 
case 2: $5/2^2 37/2\rightarrow5/2^2 39/2$, 
case 3: $3/2^2 5/2^2\rightarrow3/2^2 5/2^1 7/2^1$,
case 4: $5/2^3 9/2^3\rightarrow5/2^2 9/2^4$,
case 5: $1/2^1 3/2^2 5/2^3 7/2^4\rightarrow1/2^1 3/2^2 5/2^3 7/2^3 9/2^1$.
The value $k$ is the maximum index $2m+1$ accounted for in the expansion 
(\ref{eq:GGC_KJ}) and the error is the relative variation on the line number. 
The notation $a(-b)$ stands for $a\times10^{-b}$. 
\label{tab:line_numbering}}
\end{table}
A series of examples is provided in the table \ref{tab:line_numbering} for 
various pairs of relativistic configurations. The number of lines obtained 
with the explicit sum (\ref{eq:ntransDE}) is compared to the exact value 
as a function of the cutoff index $k_\text{max}$. To put it differently, 
the number of terms retained in Gram-Charlier-like expansion for both 
configurations is $(k_\text{max}+1)/2$. The conclusions drawn for the 
angular momentum distribution may be reiterated here. 
In the simple case $9/2^3\rightarrow9/2^2 11/2$, the three-term expansion 
involving cumulants up to the kurtosis and used in Ref.~\cite{Bauche1987} 
provides a fair approximation, which is not significantly improved by adding 
many more terms. In the special case $5/2^2 37/2\rightarrow5/2^2 39/2$, for 
which we have seen that the kurtosis-based approximation fails, the same 
conclusions hold here and a large number of terms are required in order to 
get an acceptable number of lines. In the case $3/2^2 5/2^2\rightarrow3/2^2 
5/2^1 7/2^1$, little improvement is brought by adding terms beyond the third 
($k_\text{max}=5$). Similar conclusions hold in the case $5/2^3 9/2^3
\rightarrow5/2^2 9/2^4$. One notes that a better accuracy is achieved when 
more electrons are involved. Finally in the multiple-subshell case $1/2^1 
3/2^2 5/2^3 7/2^4\rightarrow1/2^1 3/2^2 5/2^3 7/2^3 9/2^1$, one notes a 
regular increase of the accuracy provided by the expansion, the first three 
terms providing an accuracy below the percent level. Finally it is worth 
mentioning that increasing $k_\text{max}$ to very high values --- which 
depend on the pair of configurations analyzed --- one notes that the 
expansion begins to diverge, in agreement with the observation done on 
the number of levels $Q(J)$. As mentioned previously, the expansion derived 
here is asymptotic, and the larger the total number of electrons the better 
the approximation.

\section{Conclusion}
The generating function previously obtained for the distribution of the 
magnetic quantum number $M$ has been used to derive properties on the 
distribution of the total angular momentum $J$. It has been shown that in 
any single-subshell relativistic configuration this distribution can be 
efficiently computed using recurrence relations, and that even-odd 
staggering can be described for a single subshell with an even number of 
electrons. 
The analysis on the magnetic-quantum-number distribution allows us to obtain 
an expression for the $J$-distribution that generalizes the Bethe formula in 
the form of a Gram-Charlier-like expansion. 
This expansion has been tested in a series of cases and proved to be 
efficient, even in most cases when few terms are accounted for. 
However, as for the magnetic-quantum-number-distribution analysis, we 
observe the Gram-Charlier-like expansion is probably not convergent but 
of asymptotic nature. 
The formulas obtained also provide accurate approximations for the total 
number of levels in a relativistic configuration and for the number of lines 
in a transition array.

\appendix
\section{Description of the even-odd staggering in the angular-momentum 
distribution using the generating function}\label{sec:staggering}
As pointed out earlier \cite{Pain2013}, the generating function 
technique is also useful when analyzing the predominance of the levels with 
even $J$ values. 
From definition (\ref{eq:GjNz}) and the symmetry property on $Q(J)$ 
(\ref{eq:symQ}), one has for a single-subshell configuration with an 
\emph{even} number of electrons $N$
\begin{subequations}\label{eq:excess}\begin{align}\nonumber
 \mathscr{G}(j,N;-1) &= (-1)^{J_\text{max}+1}
 \left[Q(0)-Q(1)+Q(2)\cdots+(-1)^{J_\text{max}}Q(J_\text{max})\right]\\
 &\quad+(-1)^{J_\text{max}}\left[Q(-1)-Q(-2)+Q(-3)\cdots+(-1)^{J_\text{max}}
 Q(-J_\text{max}-1)\right]\\
 &= -2\mathscr{E}
 \end{align}\end{subequations}
where $\mathscr{E}$ is the even-odd excess
\begin{equation}
 \mathscr{E} = \sum_{n\ge0}Q(2n)-\sum_{n\ge0}Q(2n+1).
\end{equation}
We have used the value $J_\text{max}=N(2j+1-N)/2$ from which 
$(-1)^{J_\text{max}+1}=-1$ if $N$ is even. Therefore the evaluation of 
the excess is directly related to $\mathscr{G}(j,N;-1)$. Furthermore 
for a multi-subshell configuration, the above relation still holds while 
the product in (\ref{eq:GjNz}) is replaced by the product of the 
contribution of every subshell.
The evaluation of the excess is therefore reduced to computing the product 
(\ref{eq:GjNz}) at $z=-1$. To this respect, one notices that this quantity 
involves both at the numerator and the denominator $N/2$ factors of the 
form $(z^{2r}-1)$ which vanish at $z=-1$. The indeterminacy is removed by 
the de L'Hospital rule which amounts to replace these factors by their 
derivative. Namely one has, after eliminating factors with odd powers,
\begin{subequations}\begin{align}
 \mathscr{E} &=-\frac12\mathscr{G}(j,N;-1) = 
 \lim_{z\to-1}\prod_{r=1}^{N/2} 
\frac{z^{2j+3-2r}-1}{z^{2r}-1}\\
 &= \prod_{r=1}^{N/2} \frac{2j+3-2r}{2r} =\binom{j+1/2}{N/2}
\end{align}\end{subequations}
in agreement with Ref.~\cite{Bauche1997}. In the case where $N$ is odd, no 
simple relation like (\ref{eq:excess}) holds. However, a direct computation 
of the $Q(J)$ distribution allows one to check that the staggering on 
$Q(J)$ between $J=2n+1/2$ and $J=2n+3/2$ is much less important: one may, 
e.g., look at results presented in subsection \ref{sec:5half-filled}. The 
same formalism also applies for any multi-subshell relativistic configuration 
with an \emph{even} number of electrons. Then, Eq.~(\ref{eq:excess}) holds 
provided $\mathscr{G}(j,N;z)$ is replaced by $(z-1)$ multiplied by the 
product of the $\mathscr{F}(j_s,N_s;z)$, $\mathscr{F}$ being defined by 
Eq.~(\ref{eq:FjNz}). If every subshell has an even number of electrons, 
due to this factorization property, the excess is simply given by 
\begin{equation}
 \mathscr{E}=\prod_{s=1}^w \binom{j_s+1/2}{N_s/2}\text{ if all $N_s$ even}
\end{equation}
while $\mathscr{E}=0$ if at least one of the subshells has an odd 
occupation number. This last point is easy to verify. Let us consider a 
factor $\mathscr{F}(j_s,N_s;z)$ with $N_s$ odd. From the definition 
(\ref{eq:FjNz}), the denominator contains $(N_s-1)/2$ factors with an even 
power $z^{2r}-1$, while there are $(N_s+1)/2$ such factors in the 
numerator. When evaluating such product at $z=-1$, the result will be zero.

One may also generalize the above derivation to the formal case where $j$ 
is integer. In this case the total momentum $J$ is integer, and the excess 
can be computed whatever $N$. Noting that $J_\text{max}=N(2j+1-N)/2$ has 
the parity of $N/2$ for even $N$ and of $(2j+1-N)/2$ for odd $N$, the same 
method as above provides the excess value
\begin{equation}
  \mathscr{E} =
  \begin{cases}
    \displaystyle
    (-1)^{N/2}\binom{j}{N/2} & \text{ for even $N$} \\
    \displaystyle
    (-1)^{(2j+1-N)/2}\binom{j}{(N-1)/2} & \text{ for odd $N$}
  \end{cases}\label{eq:exc_intj}
\end{equation}
which indicates that the excess never vanishes. For a multi-subshell 
configuration, the same argument as above proves that the excess is given 
by the product of the individual contributions (\ref{eq:exc_intj}). The 
excess never vanishes in this case too.

\section{Alternative expression for the Gram-Charlier-like expansion}
\label{sec:psGC_moments}
The expression (\ref{eq:d_vs_kappa}) for the coefficients of the Gram-%
Charlier-like expansion is formally simple. However from a computational 
point of view it requires the definition of the partitions of the even 
integer $2m$ which becomes rather time-consuming for large $m$ since the 
number of such partitions increases exponentially with $m^{1/2}$ as 
shown by the Ramanujan-Hardy formula. An alternative formulation uses the 
moments of the $P(M)$ distribution 
\begin{equation}\mu_k=\sum_M M^kP(M)/\sum_MP(M).\end{equation}
Such moments can be straightforwardly derived from the expression of the 
cumulants using the well-known formula \cite{Stuart1994}
\begin{equation}
\mu_{2k}=\kappa_{2k}
  +\sum_{m=1}^{k-1}\binom{2k-1}{2m-1}\kappa_{2m}\mu_{2k-2m},
\end{equation}
which accounts here for the fact that the $P(M)$ distribution is 
symmetric, so that odd-order moments vanish. 
The cumulants for the $P(M)$ distribution are given by 
Eq.~(\ref{eq:cumul}). 
From the relation (\ref{eq:d_vs_c}) and from the standard equation 
providing the Gram-Charlier coefficients as a function of the 
moments,
\begin{equation}\label{eq:cGC_mu}
c_{2n}=\sum_{j=0}^{n}\frac{(-1)^j\mu_{2n-2j}/\sigma^{2n-2j}}{2^j j!(2n-2j)!}
\end{equation}
one gets, with the variable change $q=p+j$ so that $n-j=m-q$, and using the 
Chebyshev-Hermite polynomial expansion (\ref{eq:Hermite}),
\begin{subequations}\begin{align}
 d_{2m+1}&=\sum_{\substack{n,p,j\\n+p=m}}
 \frac{(2\sigma)^{-2p}}{(2p+1)!} 
  \frac{(-1)^j\mu_{2n-2j}/\sigma^{2n-2j}}{2^j j!(2n-2j)!} \\
  &=\sum_{\substack{n,q,j\\n+q-j=m}}
  \frac{(2\sigma)^{2j-2q}}{(2q-2j+1)!} 
  \frac{(-1)^j\mu_{2m-2q}/\sigma^{2m-2q}}{2^j j!(2m-2q)!}\\
  &=\sum_{q=0}^m \frac{2\sigma}{(2m-2q)!}\frac{\mu_{2m-2q}}{\sigma^{2m-2q}}
   \sum_{j=0}^q\frac{(-1)^j(2\sigma)^{2j-2q-1}}{2^j j!(2q+1-2j)!}\\
  &=2\sigma\sum_{q=0}^m \frac{\mu_{2m-2q}/\sigma^{2m-2q}}{(2m-2q)!(2q+1)!}
    He_{2q+1}((2\sigma)^{-1})\\
  &=\frac{2\sigma}{(2m+1)!}\sum_{q=0}^m \binom{2m+1}{2q+1}
    He_{2q+1}((2\sigma)^{-1})\frac{\mu_{2m-2q}}{\sigma^{2m-2q}}.
\end{align}\end{subequations}
This expression is almost as simple as (\ref{eq:cGC_mu}) since the 
Hermite polynomials can be computed very efficiently using simple 
recurrence relations. The first coefficients are
\begin{subequations}\begin{align}
d_3 &= \frac{1}{24\sigma^2}\\
d_5 &= \frac{\mu_4}{24\sigma^4}+\frac{1}{1920\sigma^4}-\frac{1}{8}\\
d_7 &= \frac{\mu_6}{720\sigma^6}+\frac{\mu_4}{576\sigma^6}
 -\frac{\mu_4}{48\sigma^4}+\frac{1}{322560\sigma^6}
 -\frac{1}{192\sigma^2}+\frac{1}{24}.
\end{align}\end{subequations}

\section{Gram-Charlier-like analysis for a subshell with several electrons}\label{sec:psGC_j15n7-8}
\begin{figure}[htbp]
\centering
\subfigure[Configuration $j_1=15/2, N_1=8$]{\label{fig:QJ_GC_j15n8}%
\includegraphics[scale=0.3,angle=\anglefig]{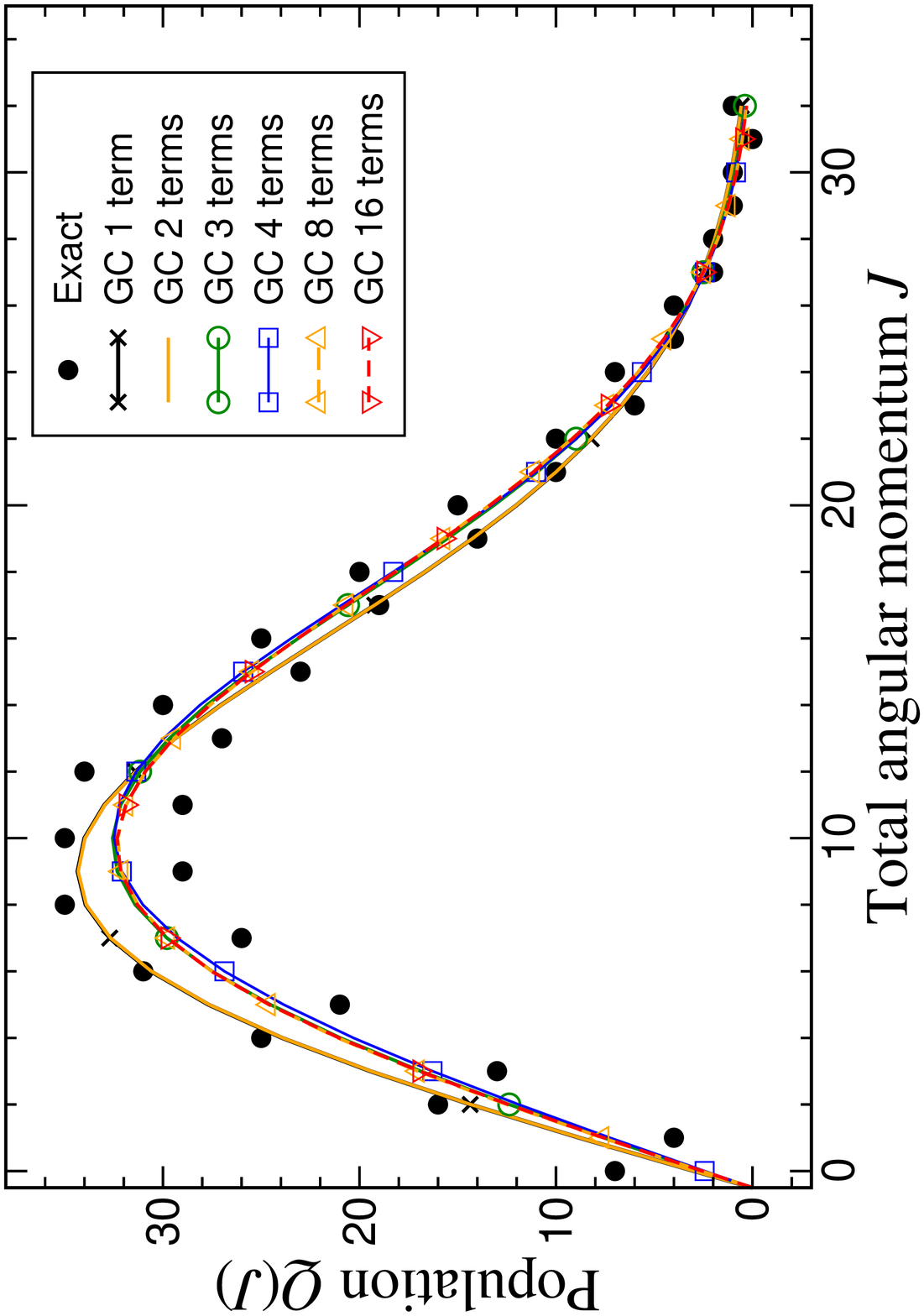}%
}
\hspace{0.1cm}
\subfigure[Configuration $j_1=15/2, N_1=7$]{\label{fig:QJ_GC_j15n7}%
\includegraphics[scale=0.3,angle=\anglefig]{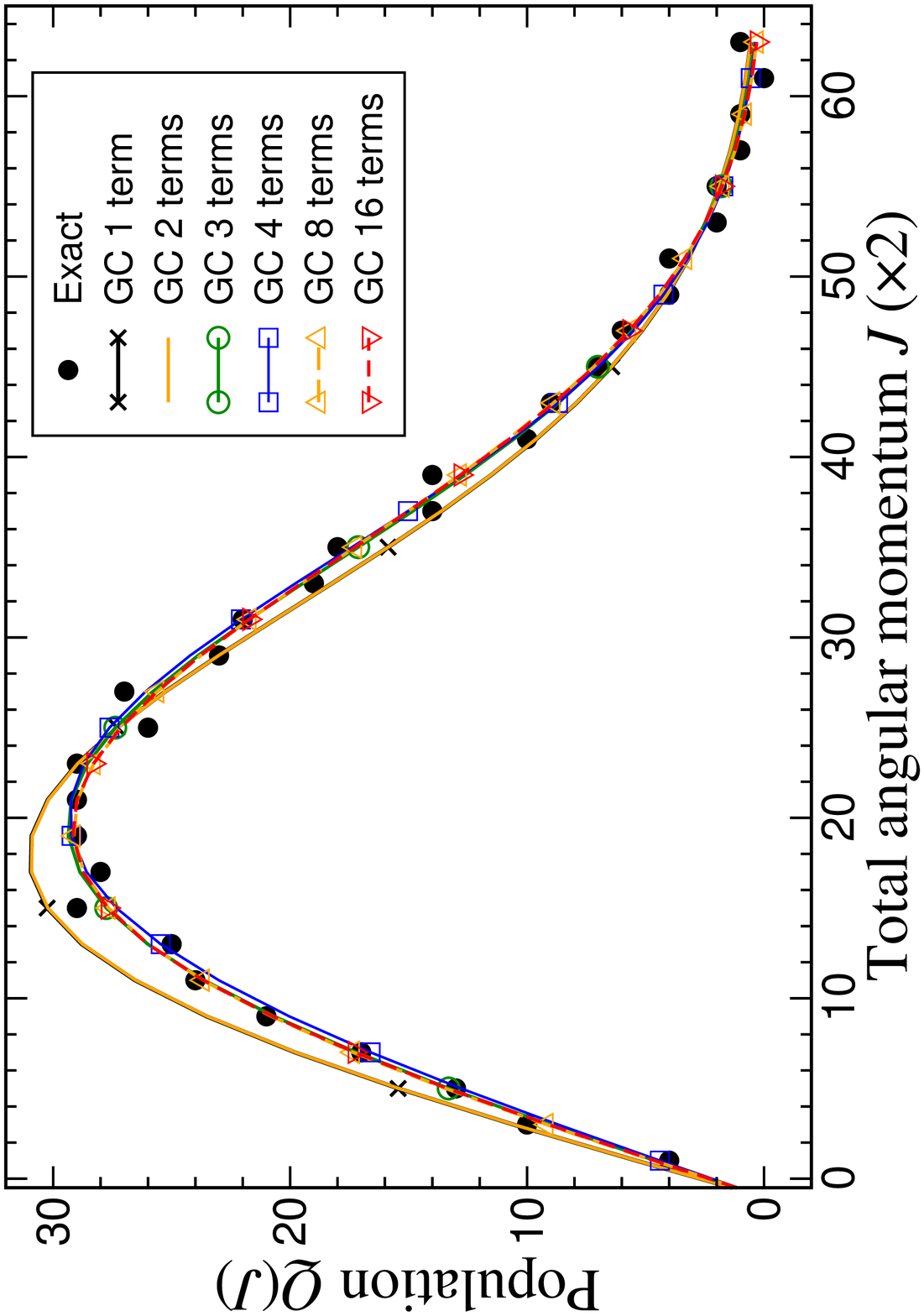}%
}
\caption{Exact and Gram-Charlier-like approximation for the total angular 
momentum distribution $Q(J)$ in the relativistic configuration with one  
subshell $j_1=15/2, N_1=8$ or $7$.\label{fig:QJ_GC_j15n7-8}}
\end{figure}

We consider in Fig.~\ref{fig:QJ_GC_j15n8} the $J$ distribution in the 
case of a single subshell $j=15/2$, $N=8$. One notices a rough agreement 
with the Gram-Charlier-like expansion, even with few terms. 
However the odd-even staggering noticed by Bauche and Coss\'e  
\cite{Bauche1997,Pain2013} and revisited in Appendix \ref{sec:staggering} is 
not reproduced even with a large number of terms. We observe that the 
expansion proposed here provides a good agreement with the distribution 
``averaged''  on even and odd cases. 

Finally Fig.~\ref{fig:QJ_GC_j15n7} presents the $J$ distribution for the 
single-subshell configuration $j=15/2$, $N=7$. The Gram-Charlier-like 
expansion provides an acceptable approximation when only 3 terms are 
accounted for, with little improvement brought by a greater number of terms. 
One notes that some staggering is observed again between the angular momenta 
$J=2n+1/2$ and $J=2n-1/2$, though this effect is smaller than the even-odd 
variation observed for an even number of electrons.

\bibliography{jrnlabbr,J-distrib}

\end{document}